\documentclass[preprint,pra,eqsecnum]{revtex4-1}
\pdfoutput=1
\usepackage{amsmath}
\usepackage{amssymb}
\usepackage{graphicx}
\usepackage{enumitem}

\newcommand{\diag}[0]{\text{diag}}

\usepackage{bm}
\renewcommand{\vec}[1]{\boldsymbol{\mathbf{#1}}}
\usepackage[font={footnotesize}]{caption}
\usepackage{hyperref}
\hypersetup{pdftex,colorlinks=true,allcolors=blue}
\usepackage{hypcap}
\begin{document}

\title{Entanglement Entropy and Mutual Information of Circular Entangling Surfaces in $2+1$-dimensional Quantum Lifshitz Model}
 
\author{Tianci Zhou}
\email{tzhou13@illinois.edu}
\affiliation{Department of Physics and Institute for Condensed Matter Theory, University of Illinois at Urbana-Champaign, 1110 West Green Street, Urbana, Illinois 61801-3080, USA}
 
\author{Xiao Chen}
\email{xchen13@illinois.edu}
\affiliation{Department of Physics and Institute for Condensed Matter Theory, University of Illinois at Urbana-Champaign, 1110 West Green Street, Urbana, Illinois 61801-3080, USA}

\author{Thomas Faulkner}
\email{tomf@illinois.edu}
\affiliation{Department of Physics and Institute for Condensed Matter Theory, University of Illinois at Urbana-Champaign, 1110 West Green Street, Urbana, Illinois 61801-3080, USA}

\author{Eduardo Fradkin}
\email{efradkin@illinois.edu}
\affiliation{Department of Physics and Institute for Condensed Matter Theory, University of Illinois at Urbana-Champaign, 1110 West Green Street, Urbana, Illinois 61801-3080, USA}

\date{\today}

\begin{abstract}
We investigate the entanglement entropy (EE) of circular entangling cuts in the 2+1-dimensional  quantum Lifshitz model, whose ground state wave function is a spatially conformal invariant  state of the Rokhsar-Kivelson type, whose weight is the Gibbs weight of 2D Euclidean free boson. We show that the finite subleading corrections of EE to the area-law term as well as the mutual information are conformal invariants and calculate them for cylinder, disk-like and spherical manifolds with various spatial cuts. The subtlety due to the boson compactification in the replica trick is carefully addressed. 
We find that in the geometry of a punctured plane with many small holes, the constant piece of EE is proportional to the number of holes, indicating the ability of entanglement to detect topological information of the configuration. Finally, we compare the mutual information of two small distant disks with Cardy's relativistic CFT scaling proposal. We find that in the quantum Lifshitz model, the mutual information also scales at long distance with a power determined by the lowest scaling dimension local operator in the theory.
\end{abstract}

\maketitle
\section{Introduction}
Entanglement entropy (EE) is a quantum information theoretical measure that can successfully capture universal properties of the many-body wave function. One celebrated example is the case of  the $1+1$-dimensional critical systems. In these systems the von Neumann entanglement entropy (EE) of a macroscopically large singly connected subsystem A  of linear size $L_A \ll L$ (where $L$ is the size of the entire system) has the leading order form $S_{\text{vN}}=\frac{c}{3}\ln\left(\frac{L_A}{\epsilon}\right)$, where $c$ is the central charge of its underlying conformal field theory (CFT), $\epsilon$ is the short-distance cutoff and $L_A \gg \epsilon$ \cite{callan1994, calabrese_entanglement_2004}. This logarithmic scaling behavior distinguishes a $1+1$-dimensional conformal field theories (CFT) from  short-range entangled states, which exhibit instead an ``area'' law, which here means constant scaling, and indicates that CFTs are long-range entangled states.

For the CFTs with spatial dimension $d>1$, von Neumann EE is expected to obey the area law $S_{vN} = \alpha \Big( \frac{L_A}{\epsilon} \Big)^{(d-1)}$ where $L_A$ is the linear size of the subsystem A. The coefficient $\alpha$ is non-universal and depends on the short-distance (UV) behavior of the model. The area law scaling behavior originates from the short-range entanglement on the boundary between A and its complement. In addition to this  non-universal leading term, several  subleading corrections are possible, some of which have been proposed to be universal \cite{casini2009, casini_towards_2011, ryu2006a, ryu2006b}.

Recently, there has been much progress in characterizing this subleading correction in $2 + 1$-dimensional CFTs for subsystems with various geometries. It was shown that in the infinite two dimensional plane, if the subsystem $A$ has the shape of a disk the subleading correction for the von Neumann EE is a finite constant (called $-F$ in the literature) \cite{jafferis2011, casini_towards_2011, casini_mutual_2015}. If, instead, the  system itself s a finite disk, this constant term is replaced by a scaling function that is constrained by the condition of strong subadditivity \cite{Lieb1973}. On the other hand, logarithmic contributions to the EE are found when the entangling region has cusp-like conical singularities on its boundary, with 
an universal coefficient representing a measure of the number of effective degrees of freedom \cite{fradkin_entanglement_2006, casini2007, Hirata2007, Kallin2014, bueno2015, Faulkner2016}. 
On the other hand, for a subsystem defined on a cylindrical section of a torus, the EE  has instead {\em finite} subleading term \cite{casini2009, chen_scaling_2014}. This finite term is shown to be scale-invariant and  depends only on the aspect ratios of the entangling region. Also, in the limit of a very short cylinder (the ``thin slice limit''), in relativistic CFTs the subleading correction is connected to the corner correction through conformal transformation \cite{bueno2015b, krempa2016}. These results are found in several different CFTs and are confirmed by the results derived from the Ryu-Takayanagi formula \cite{ryu2006a, ryu2006b}.

For the more general scale invariant systems in $2+1$ dimensions without {\em spacetime} conformal invariance, such as the quantum Lifshitz model \cite{Ardonne-2004} and the fermionic quadratic band crossing model \cite{sun-2009}, there can also be subleading corrections depending on the geometry of the subsystem \cite{fradkin_entanglement_2006, hsu_universal_2009, stephan_shannon_2009, oshikawa_boundary_2010, stephan_entanglement_2013, chen_scaling_2014}. The Quantum Lifshitz model is a compactified free boson model in $2+1$ dimensions with dynamical exponent $z = 2$, while  the fermionic quadratic band crossing model has two bands with a quadratic band touching point. In the low energy limit, this system is equivalent to a massless Dirac spinor with a quadratic dispersion and hence it also has $z = 2$. These two scale-invariant models do not have Lorentz invariance, nevertheless the subleading correction term has a similar scaling behavior as that for relativistic CFTs.

In the case of the quantum Lifshitz model of a free boson with compactification radius $2\pi R_c$, the finite subleading term of EE  consists of   a scaling function depending on certain aspect ratios determined by the subsystem geometry and a constant term determined by the compactification radius \cite{stephan_entanglement_2013}. This constant term is determined by the zero mode sector of this critical phase and indicates that EE not only measure the local geometry but also the non-local information of the total system \cite{klich2015}. This term is  similar to the constant universal correction to the EE found in topological phase in $2+1$ dimensions  \cite{kitaev_topological_2006, levin2006}albeit with a different sign.

Many previous works have demonstrated the significance of the subleading correction term $S_0$ of the EE for the quantum Lifshitz model on the cylinder geometry
 \cite{fradkin_entanglement_2006,hsu_universal_2009,stephan_shannon_2009,oshikawa_boundary_2010}. In this paper we generalize these results by investigating the structure of  the subleading corrections of EE for other entanglement surfaces. In particular, we investigate the dependence of $S_0$ on the geometry of the manifold and on the entanglement cut (the surgery). The manifolds we  consider here include the cylinder, the disk and the sphere. It turns out that on the sphere, if the subsystem is a single spherical cap, $S_0$ is only a function of compactification radius and is independent of linear size of both subsystem and the total system. For other manifolds, $S_0$ can also have a scaling function which now depends  on the aspect ratios of the entangling surface. We study this scaling function in several asymptotic limits and show that they always satisfy the strong subadditivity constraint. We also consider the geometry of a punctured plane with many holes, and  find that the constant term in $S_0$ is proportional to the number of holes of the subsystem. In addition, we study the scaling behaviors of the mutual information of two regions for both the sphere and the plane. We compare the results on these two manifolds by utilizing the spatial conformal symmetry of the wave function. We also compare our results with Cardy's  results for relativistic free field CFTs of two disjoint circles in the large separation limit \cite{cardy_results_2013}. We find that they both show similar scaling behavior although with different critical exponents.

The structure of this paper is  as follows. We first introduce the quantum Lifshitz model and present the replica method for calculating EE in section \ref{sec:q-lifshitz}. The zero mode sector is separated out to avoid the compactification issue encountered in \cite{hsu_universal_2009}.  Then we calculate the EE of disk, annulus, spherical cap(s), disjoint disks in section \ref{sec:EE_for_diff_geos}. 
We present here a calculate of the partition functions, the regularized determinant and the winding sector contribution on subsystems of these manifolds. Interesting limiting behaviors are highlighted. 
We further study the mutual information on the infinite plane and sphere in section \ref{sec:mu}. 
We summarize and conclude in section \ref{sec:summary}. The appendices are devoted to  details of the calculations and  techniques used in this paper.

\section{The Quantum Lifshitz Model, Replica Trick and Conformal Invariance}
\label{sec:q-lifshitz}

\subsection{The Quantum Lifshitz Model}
\label{sec:q-lifhitz-model}

The quantum Lifshitz model (QLM) is an effective field theory of the quantum dimer model \cite{Rokhsar1988} and its generalizations \cite{castelnovo-2005,moessner-2001,henley_relaxation_1997,Ardonne-2004, Freedman-2004, Fendley-2008}. Its Hamiltonian describes a free bose field with dynamical scaling exponent $z = 2$. The Hamiltonian of the QLM is given by
\begin{equation}
  H_0=\int d^2x \, \frac{1}{2} \Big\{ \Pi^2+   (\frac{k}{4\pi})^2  \big[{\bm \nabla}^2\phi\big]^2 \Big\} \, ,
  \label{eq:QLM-H}
\end{equation}
where $\phi(x)$ is a compactified (i.e. periodic) bosonic field, and $\Pi(x)$ is its conjugate canonical momentum. In the context of the quantum dimer model, the compactified  bose field $\phi(x)$ is obtained by coarse-graining the height variables of the dimer configurations on a bipartite 2D lattice \cite{Ardonne-2004,moessner_quantum_2011, Fradkin-2004}.
The correlations of the Rokhsar-Kivelson quantum dimer model on square lattice are described  by the choice of $k = \frac{1}{2}$ \cite{Ardonne-2004}. 

The ground state wavefunction of the QLM Hamiltonian of Eq.\eqref{eq:QLM-H} has a simple and elegant form \cite{Ardonne-2004}
\begin{align}
|\psi \rangle = \frac{1}{\sqrt{\mathcal{Z}}}\int [d\phi] e^{-\frac{1}{2} S[\phi]} |\phi \rangle,
\label{eq:Psi-QLM}
\end{align}
Here $\mathcal{Z}$ is the partition function of the free compactified boson (Gaussian) model in 2D Euclidean space,  and $S[\phi]$ is  Euclidean action of this model,
\begin{equation}
 \mathcal{Z} = \int [d\phi] e^{-S[\phi]}, \qquad S[\phi] = \frac{\kappa}{4\pi} \int d^2x \, \left({\bm \nabla}\phi \right)^2
 \label{eq:Psi-QLM2}
\end{equation}
This wavefunction of Eq.\eqref{eq:Psi-QLM} is the continuum version of the  Rokhsar-Kivelson (RK) state for the quantum dimer model \cite{Rokhsar1988}, whose amplitude for each field configuration is the Gibbs weight for the 2D classical dimer model. This model, in particular, is at a conformal quantum critical point \cite{Ardonne-2004}. 
We will exploit this property in the calculation which follows.
From now on we take $g = \frac{\kappa}{4\pi}$ and compactification radius to be $2\pi R_c$ (to be consistent with CFT conventions \cite{difrancesco_conformal_2012}).

It is important to stress that, in spite of the local form of this wave function, this theory has long-range entanglement due to the {\em compactified} nature of the field. As we will see here, consistent with earlier results on toroidal geometries \cite{stephan_shannon_2009,oshikawa_boundary_2010,Hsu-2010}, the compactified nature of the field leads to finite universal terms in several geometries that we will study here. In particular, this implies that these wave function cannot be trivially factorized on  partitions.

\subsection{Replica Trick Calculation}
\label{sec:replica-trick}

We now consider the constant term of the entanglement entropy for the RK state. The main strategy is to compute the R\'enyi entropies for a bipartition of a system into two complementary subsystems $A$ and $B$ using the  normalized RK state of Eq. \eqref{eq:Psi-QLM} and following the replica approach of Ref. \cite{fradkin_entanglement_2006} which we reproduce here.
To facilitate the derivation, we use a discrete notation for the clarity, setting the RK state to be
\begin{equation}
|\psi \rangle = \sum_{\phi} e^{- \frac{1}{2} S[\phi]} | \phi \rangle 
\label{eq:pure-state}
\end{equation}
Here $|\phi\rangle$ is a complete set of orthonormal states which are eigenstates of the field operator $\phi(x)$ of the QLM.  We will use $| a\rangle$ and $|b\rangle$ to represent two complete orthonormal basis states for regions $A$ and $B$, respectively. 

\begin{figure}[h]
\centering
\includegraphics[width=0.6\textwidth]{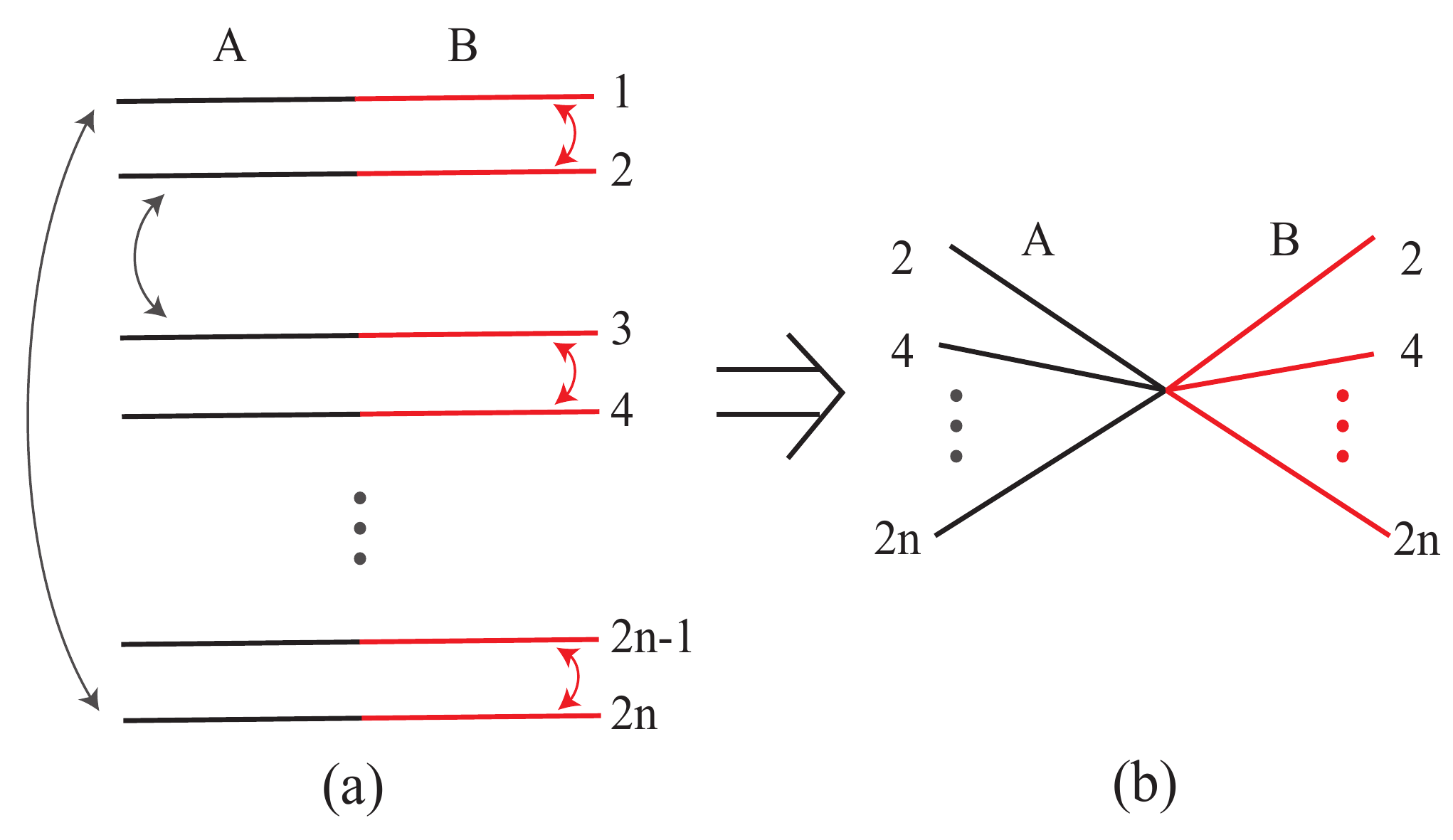}
\caption{(a) Stitching $2n$ copies in the calculation of $\text{tr}\rho_A^n$ shown in Eq.\eqref{eq:replica_trick}. (b) The book configuration of $\text{tr}\rho_A^n$ after gluing $2n$ copies together. They have the same boundary conditions along the cut.}
\label{fig:gluing}
\end{figure}

For a pure state of the combined system $\big| \psi \rangle$ (e.g. the state of Eq.\eqref{eq:pure-state}),  the density matrix is $\rho = | \psi \rangle \langle \psi |$. The reduced density matrix $\rho_A$ for subsystem $A$ is obtained by tracing over the degrees of freedom on region $B$, $\rho_A=\textrm{tr}_{B} \rho$. The trace of $\rho_A^n$ is
\begin{align}
\text{tr}(\rho_A^n) & = \sum_{b_i}\text{tr}\big[ \langle b_1 |\rho | b_2 \rangle \delta_{b_1 b_2} \cdots \langle b_{2n-1} |\rho | b_{2n} \rangle \delta_{b_{2n-1} b_{2n}} \big]\nonumber \\
&= \sum_{a_i, b_i}\langle a_1 | \langle b_1 |\rho | b_2 \rangle |a_2\rangle  \cdots \langle a_{2n-1} | \langle b_{2n-1} |\rho | b_{2n} \rangle |a_{2n}\rangle 
   \delta_{a_1 a_{2n}} \prod_{i=1}^{n-1}\delta_{a_{2i} a_{2i+1}} \prod_{i=1}^{n}\delta_{b_{2i-1} b_{2i} } \nonumber \\
&\propto \sum_{a_i,b_i} \exp\Big\{- \sum_{i=1}^{2n} \frac{1}{2}S[\phi_i] \Big\} \delta_{a_1 a_{2n}} \prod_{i=1}^{n-1}\delta_{a_{2i} a_{2i+1}} \prod_{i=1}^{n}\delta_{b_{2i-1} b_{2i} }.
\label{eq:replica_trick}
\end{align}
Here the $2n$ copies of fields $\phi_i$ are created by stitching together the states $a_i$ and $b_i$ and this process is also shown in Fig. \ref{fig:gluing} (a). The delta functions enforce the gluing condition. For example, $\phi_{2i}$ can be reproduced by taking region $A$ of $\phi_{2i+1}$ and region $B$ of $\phi_{2i-1}$. Hence we can keep only the fields of even indices, which  are independent except for the condition that they must be  equal on the entanglement cut as a result of the gluing. Therefore, the trace becomes 
\begin{equation}
 \text{tr}(\rho^n_A) = \frac{\mathcal{Z}_n(\text{equal on cut})}{\mathcal{Z}^n}, 
 \label{eq:replica}
\end{equation}
where the quantity $\mathcal{Z}_n$ in the numerator of Eq.\eqref{eq:replica}  is the partition function of $n$ copies of fields which are equal to each other on the cut. (Fig.\ref{fig:gluing} (b))

Fradkin and Moore \cite{fradkin_entanglement_2006} proposed to compute $\mathcal{Z}_n$ by performing an orthogonal transformation among the $n$ fields $\phi_i$. 
The basic idea is that since the fields $\phi_i$ (with $i=1,\ldots,n$) must take the same value on the cut $\Gamma$,  the difference of consecutive fields vanishes on the cut $\Gamma$,  while ``a center of mass'' field is unaffected by this condition. In Appendix \ref{app:topo-sum} we specify the rotation  matrix such that the resulting $n-1$ out of $n$ copies have Dirichlet boundary conditions on the cut and there is no restrictions on the remaining average (free) field. Using this argument, they  consequently obtained a simple expression for the trace in $S_n[A]$, 
\begin{equation}
S_n[A]=\text{tr}(\rho^n_A) =  \Big(\frac{\mathcal{Z}_{\text{Dirichlet}}}{\mathcal{Z}_{\text{Free}}}\Big)^{n-1}.
\end{equation}
There is, however, a subtle technical problem with this argument  \cite{oshikawa_boundary_2010,zaletel_logarithmic_2011}. The problem is that the rotation changes the compactification conditions for the fields. This problem can be addressed in the replica framework by doing a separate sum over the ``classical'' modes $\phi_{\text{cl}}(\vec{x})$ (i.e. the part of the field $\phi(x)$  that is not compact) with specified values on the cut and a sum over the contributions of the winding modes, which enforce the periodicity conditions, as was done in the supplemental material of the work of Zaletel {\it et al.} \cite{zaletel_logarithmic_2011}, and is reproduced below in Appendix \ref{app:topo-sum}.  The resulting formula, including a proper treatment of compactified boson, becomes
\begin{equation}
S_n[A]=\text{tr}(\rho^n_A) = \Big(\frac{\mathcal{Z}_{\text{Dirichlet}}}{\mathcal{Z}_{\text{Free}}}\Big)^{n-1}  W(n),
\end{equation}
where $W(n)$ is the sum over the different topological sectors of the compactified field over $m$ entanglement cuts
\begin{equation}
\label{eq:W-in-text}
W(n ) = \sum_{\vec{\phi}_{\text{cl}}|_{\rm cut} = 2 \pi R_c \vec{w}, \,\vec{w}\in \mathbb{Z}^{m(n-1)} } \exp( - g \int d^2 x \left({\bm \nabla} {\phi}_{\text{cl}} \right)^2  ).
\end{equation}
In some special cases like the annulus, the $W$ function actually factorizes into two independent $W$ functions, 
for each entanglement cut. In the general case however, winding sectors on different cuts can talk to each other, and we need to use the general formula \eqref{eq:W-in-text}.

The von Neumann entanglement entropy $S[A]$ is the analytic continuation of the R\'enyi entropies $S_n[A]$ to $n = 1$,
\begin{equation}
S[A] = - \lim_{n \to 1} \partial_n \text{tr}(\rho^n_A)   = - W(1)\ln\left(\frac{\mathcal{Z}_{\text{Dirichlet}}}{\mathcal{Z}_{\text{Free}} }\right) - W'(1). 
\end{equation}
where $\mathcal{Z}_{\text{Dirichlet}}$ and $\mathcal{Z}_{\text{Free}}$ represent the partition functions with Dirichlet and free boundary conditions on the cut respectively, both of which are path integrals of the free boson.

After noting that $W(1) = 1$ from the normalization of reduced density matrix
\begin{equation}
\text{tr}( \rho_A ) = W(1) = 1 
\end{equation}
(see Appendix \ref{app:topo-sum}) we find
\begin{align}
S[A] &=  \big[-\ln Z(A)\big] + \big[- \ln Z(B)] - \big[- \ln Z (A \cup B )\big]  - W'(1)\nonumber\\
&=\frac{1}{2}\ln \left(\frac{\det\Delta_A \det \Delta_B}{\det \Delta_{A\cup B} }\right) - W'(1)
\label{eq:S-formula}
\end{align}
where $\det \Delta_{\rm region}$ is the determinant of Laplacian operator $\Delta=- \nabla^2$ on the specified region, with Dirichlet boundary conditions. In other words, the contribution to the von Neumann EE $S[A]$ from the non-compact boson is a difference of the free energies $F[A]+F[B]-F[A \cup B]$ associated with the partition functions $Z(A)$, $Z(B)$ and $Z(A\cup B)$\cite{fradkin_entanglement_2006}, \emph{plus} the contribution from the winding sector $-W'(1)$ (which was missing in Ref. \cite{fradkin_entanglement_2006}).

For a RK state with an amplitude of the form of a Gibbs weight with a local interaction the entanglement entropy obeys the area law \cite{Papanikolaou-2007,stephan_shannon_2009}, i.e. the leading term should be proportional to the length of the cut. This result follows  from the results of  Cardy and Peschel for the free energy of a general 2D Euclidean CFT  for a system with a smooth boundary \cite{cardy_finite-size_1988}
\begin{equation}
F(A) = - \ln Z(A) = f_b|A| + f_s L - \frac{c}{6} \chi \; \ln (L/a) + \mathcal{O}(1)
\label{eq:F[A]}
\end{equation}
where $f_b$ and $f_s$ are the bulk and surface free energy density respectively, $|A|$ is the area of the system of linear length $L$, $c$ is the central charge of the 2D Euclidean CFT, $\chi$ is the Euler characteristic of the region,  and $a$ is a short-distance cutoff. As it is apparent, the bulk $f_b$ term  cancels-out in the difference in Eq. \eqref{eq:S-formula}. Fradkin and Moore showed that a $\log L_A$ subleading correction term 
cancels if the region $A$ has a smooth boundary: in this case the logarithmic terms of the free energies cancel out exactly in the computation of the EE $S[A]$ since the Euler characteristic $\chi$ does not change \cite{hsu_universal_2009}. On the other hand, for a region $A$ with a non-smooth boundary (i.e. boundary with a cusp or corner) there is such a logarithmic term \cite{fradkin_entanglement_2006,zaletel_logarithmic_2011}.
 Therefore,  with a proper regularization scheme, Eqs. \eqref{eq:S-formula} and \eqref{eq:F[A]} show that there is a universal sub-leading correction to the EE that can be extracted. Here we choose to use the $\zeta$ function method to regularize the determinants entering  in Eq.\eqref{eq:S-formula}, see for example \cite{elizalde_zeta_1994} for a detailed account of this method.

\subsection{Conformal invariance, the constant part of the EE, and mutual information}
\label{sec:conformal-constant}

The amplitude for a field configuration of the RK state wavefunction, given in Eqs.\eqref{eq:Psi-QLM} and \eqref{eq:Psi-QLM2}, is the same as (1/2) the Gibbs weight of the 2D free Euclidean compactified boson,  which is a 2D conformal field theory. Since the scaling dimension of free boson is zero,  the RK wavefunction is invariant under  (global) spatial conformal transformations. Then, one would expect the EE to be a conformal invariant as well. However, in field theory EE is only well defined with an explicit short distance cutoff. In fact, in the regime $ \epsilon \ll L_A \ll L$ (where $\epsilon$ is a short-distance cutoff, $L_A$ is the linear size of the observed region and $L$ is the linear size of the entire system) the EE of the RK state consists of a cutoff-dependent (and hence non-universal) area (circumference) law term, which arises from the short-range entanglement of the wave function, and a constant piece $S_0$, specifically, 
\begin{equation}
S = \alpha \frac{l_{\text{cut}}}{\epsilon} + S_0.
\end{equation}
where $l_{\rm cut} \propto L_A$.
The cutoff will change under a conformal transformation, while the cutoff independent term $S_0$ will remain invariant. Therefore, in this theory it is $S_0$ instead of the whole $S$ that is conformal invariant: the constant term, $S_0$, represents the long-range entanglement/correlations encoded in the wave function.

So far we have focused on the von Neumann EE. We will also be interested in a related measure of correlations, known as the {\em mutual information}, which is defined as follows. Let us now denote by $A$ and $B$  two {\em disconnected} subsystems of the total system. The mutual information of $A$ and $B$ is   defined to be
\begin{equation}
I[A, B] = S[A] + S[B] - S[A\cup B] 
\label{eq:mutual}
\end{equation}
where $S[A]$, $S[B]$ and $S[A \cup B]$ are the von Neumann EEs of the two regions and of their union. Clearly the ``area law'' terms will
exactly canceled. Consequently, the mutual information should also be a conformally invariant. 
 
\section{EE on cylinders, disks and spheres}
\label{sec:EE_for_diff_geos}

\subsection{Cylindrical Geometry}
\label{sec:cylinder}

\begin{figure}[h]
\centering
\includegraphics[scale=.8]{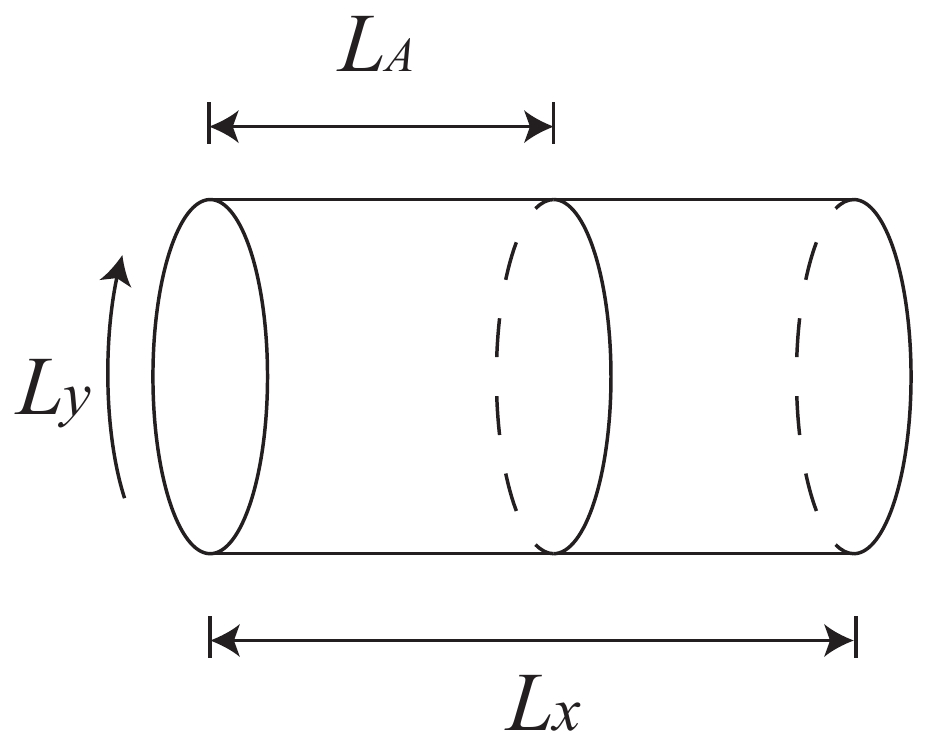}
\caption{Cylinder configuration of the system}
\label{fig:cyld}
\end{figure}

Although thoroughly studied by many authors \cite{fradkin_entanglement_2006,hsu_universal_2009,stephan_entanglement_2013,chen_scaling_2014,zaletel_logarithmic_2011}, in this section we briefly review EE for a system on a cylinder to verify the validity of the replica formula of Eq. \eqref{eq:S-formula}. 
As shown in Fig. \ref{fig:cyld}, the cylinder is cut into two cylinders, where subsystem $A$ is the one on the left. Using the results of the calculation of the required determinants summarized in Appendix \ref{app:det-cyl-ann}, we find that the subleading term $S_0$ to the (von Neumann) EE is 
\begin{align}
S_0&= \frac{1}{2} \ln \Big|\frac{2 u\tau \eta^2( 2 u\tau ) \times  2 (1 - u)\tau \eta^2( 2\tau - 2u\tau)   }{2 \tau \eta^2 ( 2 \tau ) } \Big| - W'(1)\nonumber \\
&= \ln \Big|\frac{\eta( 2 u\tau ) \eta( 2\tau - 2u\tau)   }{\eta ( 2 \tau ) } \Big|  + \frac{1}{2} \ln 2u( 1- u) |\tau|  - W'(1)
\end{align}
where $u = \frac{L_A}{L_x}$ is the aspect ratio of the observed region,  $\tau = i\frac{L_x}{L_y}$ is the complex aspect ratio of whole cylinder, and $\eta(z)$ is the Dedekind eta function. 

When subsystem $A$ is half of the infinite cylinder, i.e. $u = \frac{1}{2}, |\tau|\rightarrow \infty$,  the contribution of the winding sector derived in Appendix \ref{app:topo-sum}, leads to the result
\begin{equation}
\frac{1}{2} \ln 2u( 1- u) |\tau|  - W'(1) =  \ln \sqrt{8\pi g} R_c - \frac{1}{2}
\end{equation}
Using that $\eta( \tau) \sim \exp( -\frac{\pi}{12}|\tau| )$ in the $\tau \rightarrow \infty $ limit
\begin{equation}
\lim_{\tau\to\infty}\ln \Big|\frac{\eta( 2 u\tau ) \eta( 2\tau - 2u\tau)   }{\eta ( 2 \tau ) } \Big| = 0.
\end{equation}
we obtain that the finite term $S_0$ reaches the asymptotic limit
\begin{equation}
S_0  = \ln \sqrt{8\pi g} R_c - \frac{1}{2},
\end{equation}
Conversely, if the observed region is a thin stripe  with $u \rightarrow 0$ and $|\tau| \rightarrow \infty$,   the leading term in $S_0$ reduces to $\ln \eta( 2u \tau )$. The small parameter expansion of eta function can be done by means of a modular transformation 
\begin{equation}
  \eta( \tau) = \frac{1}{\sqrt{-i \tau}} \eta( -\frac{1}{\tau} ).
\end{equation}
Hence
\begin{align}
\eta( 2u \tau ) = \frac{1}{\sqrt{2u |\tau|}}  q^{\frac{1}{24}}\prod_{n=1}^{\infty} ( 1- q^n ) \xrightarrow{ \tau \to \infty }  \frac{1}{\sqrt{2u |\tau|}}  \exp\left( - \frac{\pi}{24 u | \tau|} \right)
\end{align}
where we used that
\begin{equation}
q = \exp\left(- \frac{\pi}{u | \tau|} \right)
\end{equation}
We therefore find that in this limit  the scaling function becomes
\begin{equation}
\label{eq:squ-1-over-alpha}
S_0 \simeq  \ln \eta( 2u \tau ) = -\frac{\pi}{24|\tau|} \frac{1}{u} .
\end{equation}
The universal $\frac{1}{u}$ dependence is also reported in Refs. \cite{stephan_entanglement_2013,chen_scaling_2014}. 

\subsection{Disk Geometry}
\label{sec:circular}

We will now discuss the case of a system on a disk with Dirichlet boundary conditions  
and an additional circular entangling cut interior to the disk. This situation is analytically tractable and the result demonstrates several important generic features about the RK wavefunctions. The result of the EE for a disk has been calculated by Ref.\cite{fradkin_entanglement_2006}, which did not treat the compactified boson correctly.
Here we derive  results that treat the compactified boson correctly, show the dependence of the finite terms in the entanglement entropy on the aspect ratio of the disk,  and give a detailed analysis of the results in the asymptotic regimes $u \to 0$ and $u \to \infty$.
This is also a check for our further study for a spherical configuration and for a plane with more punctured holes, both of which will be discussed below.

\subsubsection{Disk with Dirichlet Boundary Conditions}
\label{sec:disk}

\begin{figure}[hbt]
\centering
\includegraphics[scale=1.8]{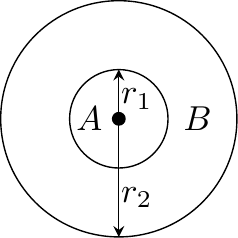}
\caption{Disk of Dirichlet boundary with radius $r_2$. The subsystem A is a concentric disk with radius $r_1$.}
\label{fig:disk-conf}
\end{figure}

We consider a subsystem $A$ which is a circle with radius $r_1$ inside a larger concentric disk with radius $r_2$, and impose Dirichlet boundary conditions on the outer boundary, see Fig. \ref{fig:disk-conf}. Its EE is given by Eq. \eqref{eq:S-formula}. Using the determinants for a disk and an annulus (see Appendices \ref{app:sph-cap} and \ref{app:det-cyl-ann}), we have 
\begin{equation}
\frac{\det \Delta_A }{\det \Delta_{A\cup B}} = \Big(\frac{r_1}{r_2}\Big)^{-\frac{1}{3}},\quad 
\det \Delta_B  =  \frac{1}{\pi} \Big(\frac{r_1}{r_2}\Big)^{\frac{1}{3}} \ln \Big(\frac{r_2}{r_1}\Big) \prod_{n\ge 0} \Big[ 1 - \big(\frac{r_1}{r_2}\big)^{2n} \Big]^2 .
\end{equation}\\
We now consider various limits for $r_1$ and $r_2$.

\paragraph{$r_2\gg r_1$\newline}

In Appendices \ref{app:topo-sum} and \ref{app:W-ann} we show that the appropriate winding function $W(n)$ is a multidimensional theta function and evaluate its scaling behaviors in various limits. We quote the approximation for the regime $r_2 \gg r_1$, 
\begin{equation}
-W'(1) \simeq - \frac{1}{2} \ln \ln \frac{r_2}{r_1}  + \ln \sqrt{8\pi^2 g} R_c - \frac{1}{2} - 2\Big( \frac{r_1}{r_2}\Big)^{\frac{1}{4\pi g R_c^2}}
\end{equation}
Therefore, we find the subleading correction to EE in this case is,
\begin{align}
S_0   &\simeq \frac{1}{2} \ln \bigg[\frac{1}{\pi} \ln \left(\frac{r_2}{r_1}\right) \prod_{n> 0} \big[ 1 - \big(\frac{r_1}{r_2}\big)^{2n} \big]^2\bigg] - \frac{1}{2} \ln \ln \frac{r_2}{r_1}  + \ln \sqrt{8\pi^2 g} R_c - \frac{1}{2}- 2\Big( \frac{r_1}{r_2}\Big)^{\frac{1}{4\pi g R_c^2}}  \nonumber \\
&= \ln \left(\sqrt{8 \pi g }  R_c\right) - \frac{1}{2}- \left(\frac{r_1}{r_2}\right)^{2} - 2\Big( \frac{r_1}{r_2}\Big)^{\frac{1}{4\pi g R_c^2}}-\mathcal{O}\left( \big(\frac{r_1}{r_2}\big)^{x} \right)
\label{eq:disk_disk}
\end{align}
In the later expansion of $S_0$ we have kept respectively a constant term depending on the compactification radius $R_c$ and two sub-leading powers of $r_1 / r_2$, which go to zero in the limit $r_1/r_2\to 0$. The terms we have dropped have higher powers than $2$ and $\frac{1}{4\pi gR_c^2}$ and can thus be neglected int this limit. Also notice that the term of the form $\ln [\frac{1}{\pi} \ln (\frac{r_2}{r_1})]$, which is present for a non-compactified boson \cite{fradkin_entanglement_2006}, cancels against a contribution from the winding modes in the compactified case.

For the infinite plane, $r_2 \rightarrow \infty$, $S_0$ reduces to a finite constant
\begin{equation}
S_0(r_1 , r_2 = \infty)  = \ln \sqrt{8 \pi g }  R_c - \frac{1}{2}
\end{equation}
that is independent of the size of the subsystem. On an infinite plane, circles of different radii are related by global conformal invariance. The constancy is thus a manifestation of the conformal invariance we argued before. The $-\frac{1}{2}$ missed in \cite{hsu_universal_2009} was also derived using boundary CFT methods in \cite{oshikawa_boundary_2010} and reported in numerical calculation in \cite{stephan_shannon_2009}. 

We compare the results with EE of a disk of a 2+1d CFT system, where the Hamiltonian rather than the wavefunction is conformal invariant. For 2+1d CFTs, the disk EE on an {\em infinite plane} has a subleading correction, which is a finite constant related to the regulated free energy $F$ on $S^3$\cite{casini_towards_2011,hirata_ads/cft_2007,nakaguchi_entanglement_2015}. This constant piece $F$ is universal and decreases along the RG flow \cite{Liu:2012eea,casini-huerta-2012}. 
The constant piece in our model however has a different origin: it comes from the zero mode sector of the compact boson.

The additional scaling function in Eq.\eqref{eq:disk_disk} gives finite size corrections to the constant part. From purely dimensional ground, the system has only two length scales $r_1$ and $r_2$ in a regularized theory. $S_0$ as a dimensionless cutoff independent quantity should be a function only of $\frac{r_1}{r_2}$. This scaling function in the limit $r_2 \gg r_1$ is much smaller than the constant term in $S_0$ but will become dominant in the calculation of mutual information. We will come back to this point later.\\

\paragraph{$r_2\sim r_1$\\}
On the other hand, when $r_2 \sim r_1$, the suitable small parameter is the modular parameter $\tau$ defined as
\begin{equation}
 \exp( -2\pi |\tau|) = \left(\frac{r_1}{r_2}\right)^2.
\end{equation}
In this limit $|\tau| \rightarrow 0 $, we have
\begin{equation}
W(n) = 1 + \mathcal{O}\left(\exp\left( - \frac{8 \pi^3 g R_c^2}{|\tau|}\right)\right)
\end{equation}
Up to an exponentially small error,  for $r_2 \sim r_1$ the constant term $S_0$ becomes
\begin{align}
S_0  &=  \frac{1}{2} \ln \left[ \frac{1}{\pi}\ln \Big(\frac{r_2}{r_1}\Big) \prod_{n\ge 0} \left[ 1 - \left(\frac{r_1}{r_2}\right)^{2n} \right]^2\right]\\
&\simeq -\frac{\pi}{12|\tau|}  = - \frac{\pi^2 }{12 \ln \left(\frac{r_2}{r_1} \right)}\simeq -\frac{\pi}{24} \left(\frac{2\pi r_1}{r_2-r_1}\right)=-\frac{\pi}{24} \left(\frac{l_{\text{cut}}}{r_2-r_1}\right)
\end{align}
which diverges linearly in the  ratio, $\frac{r_1}{r_2-r_1}$,  as in the thin-slice limit of a cylinder, Eq.\eqref{eq:squ-1-over-alpha}. As $r_1 \rightarrow r_2$, the degrees of freedom in subsystem $B$ is getting smaller and smaller; while the length of the cut does not change much, the subleading $S_0$ must be negative and decreasing to reduce the total EE.

In general, for system with {\em finite radius} $r_2$ (Fig. \ref{fig:disk-conf}), the EE of a disk for RK state has the form
\begin{equation}
S_{0} = -f\left(\frac{r_2}{r_1},R_c\right)
\end{equation}
where $f(\frac{r_2}{r_1},R_c)$ is a function depending on the ratio $r_2/r_1$ and $R_c$.  By applying strong subadditivity to the annulus configuration (see \cite{hirata_ads/cft_2007} or \cite{nakaguchi_entanglement_2015} for a detailed derivation), $f$ must be a monotonically decreasing and convex function of $r_2/r_1$. Our function $-f(\frac{r_2}{r_1},R_c)$ satisfies this requirement in both $r_2\gg r_1$ and $r_2\sim r_1$ limits.

\subsubsection{Annulus with Dirichlet Boundary Conditions}
\label{sec:annulus}

\begin{figure}[hbt]
\centering
\includegraphics[scale=0.8]{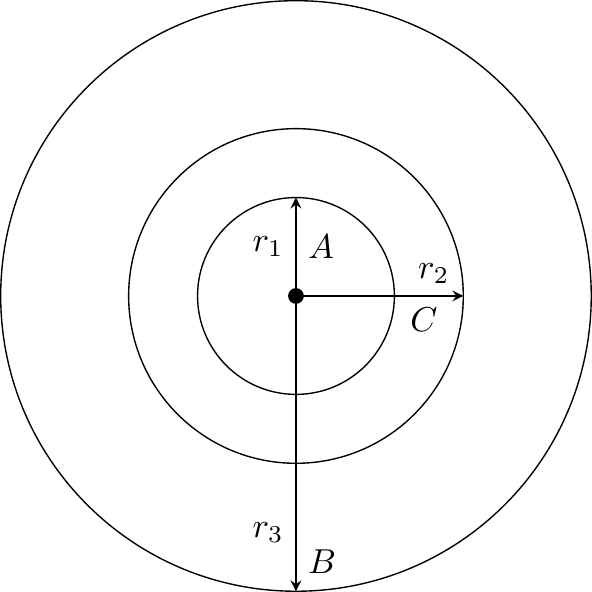}
\caption{Annulus configuration with Dirichlet boundary. The radii of $A, C$ and $B$ are $r_1, r_2$ and $r_3$ respectively. We assume $r_3\to\infty$ and $r_3 \gg r_1, r_2$.}
\label{fig:ann-conf}
\end{figure}

We now consider the case of three concentrical circles with the largest one  with radii  $r_3>r_2>r_1$, see Fig.\ref{fig:ann-conf}. In this geometry, region $A$ is the disk of radius $r_1$,  $C$ is the annular region comprised between $r_2>r>r_1$, and $B$ is the outer annular region, $r_3>r>r_2$.  We will assume Dirichlet boundary conditions on the outer circle of radius $r_3$. 

In the limit $r_3 \rightarrow \infty$, we consider the EE of the annulus $C$ in the limits of $r_1 \sim r_2$ (thin annulus) and $r_2 \gg r_1$ (thick annulus). Upon generalizing   Eq. \eqref{eq:S-formula} to this geometry, the subleading term of the EE now becomes:
\begin{align}
  S_0 &= \frac{1}{2}\ln\left[\frac{\det \Delta_A \det \Delta_B \det \Delta_C}{\det \Delta_{A\cup B \cup C} }\right] - W'(1)\nonumber\\
&= \frac{1}{2} \ln\left[\frac{\det \Delta_A \det \Delta_B \det \Delta_C}{\det \Delta_{A\cup B \cup C} }\right] - W'_{12}(1)W_{23}(1) - W_{12}(1)W'_{23}(1)\nonumber\\
&= \frac{1}{2} \ln\left[\frac{\det \Delta_A \det \Delta_B \det \Delta_C}{\det \Delta_{A\cup B \cup C} }\right] - W'_{12}(1) -W'_{23}(1)
\end{align}
Here $W_{12}$ and $W_{23}$, defined and computed in Appendix \ref{app:W-ann},  account for the contributions of the winding modes for the annular regions $C$ and $B$, respectively. In principle, one should calculate $W(n)$ function for the two sets of winding vectors on the two entanglement cuts. However, the classical modes (see definitions in Appendix \ref{app:topo-sum}) only depend on \emph{relative} winding numbers on the two edges of the annulus. Therefore the winding function effectively factorize into two independent function for region $C$ and $B$, i.e., $W(n)=W_{12}(n)W_{23}(n)$,

The calculation of the determinants is very similar to that   for the disk, except that now there are   more parameters. The result is
\begin{equation}
\frac{\det \Delta_A \det \Delta_B \det \Delta_C}{\det \Delta_{A \cup B \cup C}} = \frac{1}{\pi^2} \ln \left( \frac{r_2}{r_1}\right) \ln \left(\frac{r_3}{r_2}\right)\; \prod_{n > 0} \Big(\Big[ 1 - \big(\frac{r_1}{r_2}\big)^{2n} \Big]^2 \Big[ 1 - \big(\frac{r_2}{r_3}\big)^{2n} \Big]^2\Big)
\end{equation}
For the winding modes sector, since we require $r_3\gg r_2$ and $r_3\to \infty$,  in this limit, we have 
\begin{equation}
W_{23}'(1) \simeq \frac{1}{2} \ln \left(\frac{\ln \frac{r_3}{r_2} }{8\pi^2 g R_c^2 }\right) + \frac{1}{2}
\end{equation}\\

\paragraph{Thick annulus with $r_2 \gg r_1$\\}
For a thick annulus with $r_2 \gg r_1$, we also obtain
\begin{equation}
W_{12}'(1) \simeq \frac{1}{2} \ln \left(\frac{\ln \frac{r_2}{r_1} }{8\pi^2 g R_c^2 }\right) + \frac{1}{2}+2\left(\frac{r_1}{r_2}\right)^{\frac{1}{4\pi gR_c^2}}, 
\end{equation}
We then find that in the thick annulus regime the constant term in the EE for region $C$ becomes
\begin{align}
S_0^{\text{thick}} &\simeq 2\Big[ \ln(\sqrt{8\pi g} R_c)- \frac{1}{2}\Big]+\ln \bigg\{\prod_{n > 0} \big[1 - \big(\frac{r_1}{r_2}\big)^{2n}\big]  \bigg\}-2\left(\frac{r_1}{r_2}\right)^{\frac{1}{4\pi gR_c^2}}\nonumber\\
&= 2\Big[ \ln(\sqrt{8\pi g} R_c)- \frac{1}{2}\Big]-\mathcal{O}\left(\frac{r_1}{r_2}\right).
\end{align}
where on the second line, we neglect the term that vanishes in the $r_2 \gg r_1$ limit.
The coefficient $2$ is coming from the two sets of winding modes on the two entanglement cuts. This result can be generalized to the plane with multiple holes and will be discussed in section \ref{multi_hole}.\\

\paragraph{Thin annulus with $r_2\sim r_1$\\}
Conversely, when $r_2 \sim r_1$, we now find
\begin{equation}
W_{12}'(1) \simeq 1, 
\end{equation}
Hence,  the subleading term of the EE in the thin annulus limit for region $C$ is
\begin{align}
S_0^{\text{thin}}   &\simeq \ln(\sqrt{8\pi g} R_c)- \frac{1}{2} +  \frac{1}{2} \ln \bigg[ \frac{1}{\pi}\Big(\ln \frac{r_2}{r_1}\Big) \prod_{n > 0} \big( 1 - \big(\frac{r_1}{r_2}\big)^{2n} \big)^2\bigg]\nonumber \\  
&\simeq -\frac{\pi}{24} \left(\frac{2\pi r_1}{r_2-r_1}\right) + \ln(\sqrt{8\pi g} R_c)- \frac{1}{2}.
\end{align}
where the leading term is scaling function of $r_2/r_1$ and will diverge as $r_2/r_1\to 1$ as in the thin-slice limit of a cylinder, Eq.\eqref{eq:squ-1-over-alpha}. The second term is a constant and is coming from the winding function $W_{23}(n)$. The result of $S_0$ enables us to investigate the mutual information of two spherical caps in various limits and we will discuss them in section \ref{sec:mu}.

\subsection{Spherical Geometry}
\label{sec:sphere}

Let us now put the wavefunction on a sphere. The spherical geometry introduces the radius of sphere as a new length scale. The free energy in terms of the metric $g$ of the manifold (the sphere in this case) is an effective action
\begin{equation}
F[g] = - \ln \bigg\{\int [d\phi] e^{-S[\phi,g]} \bigg\} = \frac{1}{2} \ln \det \Delta[g]. 
\end{equation}
Furthermore the difference between $F[g]$ of curved space with metric $g_{ab}$ and $F[\delta_{ab}]$ of flat space is the trace anomaly of the stress tensor \cite{difrancesco_conformal_2012}. EE is given by the following linear combination  of these free energies
\begin{equation}
S = F[g,A] + F[g,B] - F[g, A \cup B ] - W'(1) 
\end{equation}
where $F[g,A]$ is the free energy of region $A$ of the sphere (with metric tensor $g$), etc.
Although the free energies themselves are not, both their difference(see Ref. \cite{weisberger_conformal_1987}, reproduced in Appendix \ref{app:conformal-flow}) and winding function $W$ are conformally invariant.  This implies the conformal invariance of the constant term of the EE. Therefore, we expect that the radius of curvature will not appear in the EE because a conformal map can change it arbitrarily. 

A global conformal map takes circles to circles on the sphere. Hence one would anticipate that the constant term of the EE of a single spherical cap should be independent of the angular opening of the cap. 
In addition, just as for the case of annulus on a plane, the only sensible parameter that will enter in the expressions of interest is the cross ratio for two spherical caps. 

Contrary to the disk configurations we considered before, the sphere has no boundary.
This justifies both $|\phi\rangle$ and $|\phi + \text{const} \rangle$ as plausible states. In order to remove the redundancy due to compactification, we identify the states
\begin{equation}
|\phi \rangle \equiv | \phi + 2\pi R_c \rangle
\end{equation}
in the Hilbert space or, equivalently, we constrain the range of the constant mode of $\phi$ to be in the interval $[0,2\pi R_c]$. 
For the path integral involved in the computation of the partition function, we expand the field $\phi$ as a linear combination of the eigenfunctions $\psi_j$ of the operator $-\frac{g}{\pi}\nabla^2$
\begin{equation}
\phi = \sum_{j=0}^{\infty} c_j \psi_j
\end{equation}
such that
\begin{equation}
\mathcal{Z} = \int [d\phi] e^{-\pi  \frac{g}{\pi} \int d^2 x \left({\bm \nabla}  \phi \right)^2 } = \int dc_0 \prod_{j=1}^{\infty} \int dc_j e^{-\pi \lambda_j c_j^2} =  \prod_{j=1}^{\infty} \lambda_j^{-\frac{1}{2}} \int dc_0
\end{equation}
where $\lambda_j$ are the eigenvalues. 
Particular attention should be paid to the first eigenfunction $\psi_0 = \sqrt{\frac{\pi}{gA}}$ whose eigenvalue is $\lambda_0 = 0$. We only integrate over the valid range of the coefficient $c_0 \in [0, 2\pi R_c \sqrt{\frac{gA}{\pi}}]$. Therefore, the partition function on sphere becomes
\begin{equation}
\mathcal{Z} _{\text{sphere}}= 2\pi R_c \sqrt{\frac{gA}{\pi}} \left(\det \Delta_{\rm sphere} \right)^{-1/2}
\end{equation}
where the determinant involves only the non-zero modes.
Notice that in principle we should use $\det \left(\frac{g}{\pi}\Delta_{\rm sphere}\right)$ in the expression, however this just effectively change the radius of sphere, which we have argued has no influence on EE.

In contrast to the disk case that we considered before, the mode expansion of the $\phi$ field on sphere has a constant mode which does not exist on open regions with Dirichlet boundary conditions. As a result, we obtain the following  modified formula from the replica trick method (details can be found in Appendix \ref{app:rep_formula_sph}),
\begin{equation}
\text{tr}(\rho_A^n) = \sqrt{n} \Big(\frac{\mathcal{Z}_{\rm Dirichlet}}{\mathcal{Z}_{\rm Free} } \Big)^{n-1}  W( n)
\end{equation}
where the factor $\sqrt{n}W(n)$ is the partition function for the zero-mode sector.  Here, the factor of $\sqrt{n}$ is coming from the rescaling of the compactification radius and will contribute with a term of  $-1/2$ in the constant term of the EE, $S_0$.  $W(n)$ function is a little bit different from above for the cases that we are interested in before. Now it is given by
\begin{equation}
W(n) = \left\lbrace
       \begin{aligned}
         &1 & \quad (\text{single entanglement cut}) \\
         &\sum_{\vec{w}\in \mathbb{Z}^{n-1}} \exp( - g \int d^2x \nabla \phi_{\rm cl} \cdot \nabla \phi_{\rm cl } ) & \quad (\text{two entanglement cuts}) \\
       \end{aligned} \right. 
\end{equation}
The free compact partition function consists of a determinant part and the integration of zero mode. Following the calculation in Appendix \ref{app:rep_formula_sph}, we have
\begin{equation}
\mathcal{Z}_{\rm Free}  =  2\pi R_c \sqrt{\frac{gA}{\pi}}\det{}^{-\frac{1}{2}} \Delta_{\rm sphere} 
\end{equation}
Therefore, by performing the analytic continuation for $\mbox{tr}(\rho_A^n)$, we obtain the subleading correction term $S_0$ in the EE we are interested in to be given by
\begin{equation}
S_0[A] = \frac{1}{2}\ln \left(\frac{\det\Delta_A \det \Delta_B}{ \frac{1}{A} \det \Delta_{A\cup B} }\right) + \ln\Big( \sqrt{4 \pi g} R_c \Big)- \frac{1}{2} -  W'(1)
\end{equation}

We will now consider two specific cases.

\subsubsection{Single Spherical Cap}
\label{sec:spherical-cap}

\begin{figure}[hbt]
\centering
\includegraphics[width=0.25\textwidth]{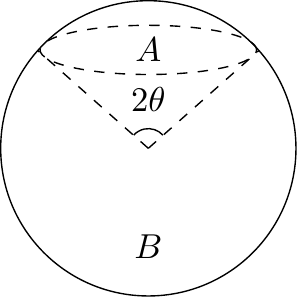}
\caption{Single spherical cap of angular opening $2\theta$.}
\label{fig:single_cap}
\end{figure}

We will now consider the case in which the subsystem $A$  is a single spherical cap of angular opening $2\theta$, as shown in Fig. \ref{fig:single_cap}. The needed determinants are found in the literature \cite{weisberger_conformal_1987, dowker_functional_1994, dowker_further_1995}. We also summarize the calculation of the determinant of the Laplacian operator for  a single spherical cap of colatitude  $\theta$ in Appendix \ref{app:sph-cap} . The result is
\begin{equation}
\frac{1}{2} \ln \det \Delta(\theta)   =  \frac{1}{2} \ln \det \Delta_{\rm hemisphere} - \frac{1}{3}\cos \theta - \frac{1}{6} \ln \tan \frac{\theta}{2} 
\end{equation}
where $\det \Delta_{\rm hemisphere}$ is the determinant of the  Laplacian for a hemisphere of unit radius. 
For the complementary spherical cap (i.e. region $B$), we set $\theta \rightarrow \pi - \theta$ and get,
\begin{equation}
 \det \Delta( \theta ) \det \Delta( \pi - \theta ) = \big|\det \Delta_{\rm hemisphere}\big|^2 .
\end{equation}

Using the exact results reported in Refs. \cite{weisberger_conformal_1987, dowker_functional_1994, dowker_further_1995} for a sphere and a hemisphere (with unit radii), 
\begin{equation}
\begin{aligned}
\det \Delta_{\rm hemisphere} &= \exp\left( - 2 \zeta'( -1) - \frac{1}{2} \ln 2\pi + \frac{1}{4} \right) \\
\det \Delta_{\rm sphere} &= \exp\left( - 4 \zeta'( -1) + \frac{1}{2} \right),
\end{aligned}
\end{equation}
we obtain the simple result
\begin{equation}
S_0 =  \ln \left(\sqrt{8\pi g} R_c \right) - \frac{1}{2}
\end{equation}
which is a constant and is indeed independent of $\theta$, consistent with the  requirement of conformal invariance.

\subsubsection{Two Spherical Caps}
\label{sec:two_cap}

\begin{figure}[h]
\begin{center}
\includegraphics[scale = 1.0]{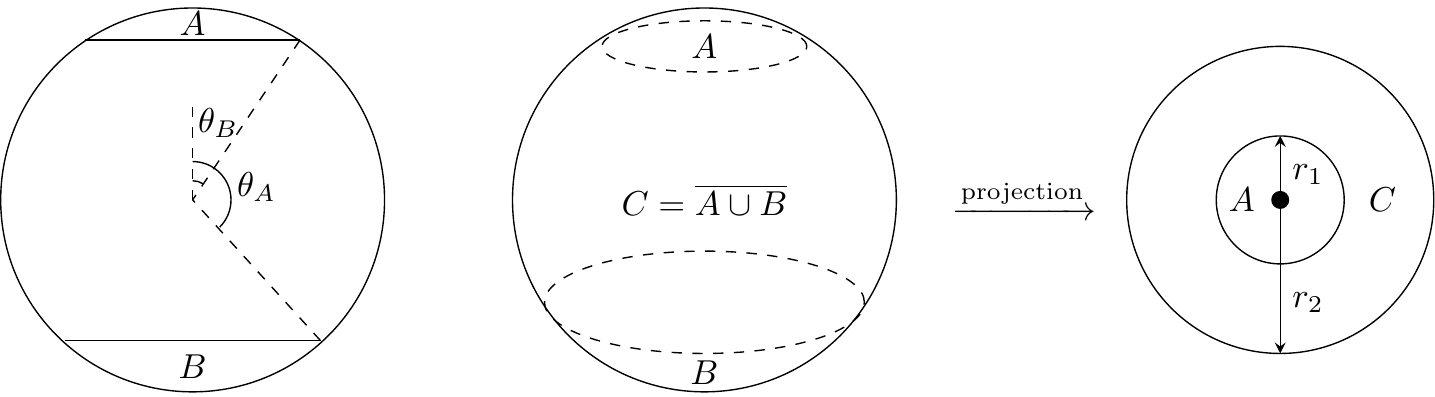}
\end{center}
\caption{Two concentrical spherical caps on the sphere. The two entanglement cuts are lines of  constant latitude. The leftmost figure is the side view of the central figure where the opening angles are specified. Regions $A$ and $B$ occupy the North and the South poles. The rightmost figure is the  stereographic  projection (from the South Pole) of the central figure. In terms of  standard spherical coordinates, the stereographic map is $z = \tan \left(\frac{\theta}{2}\right) e^{i\phi}$. So the projected variables are $r_1 = \tan \frac{\theta_A}{2}$, $r_2 = \tan \frac{\theta_B}{2}$. However, it is actually more convenient to use the projected variables rather than those angles.}
\label{fig:sphere-stripe}
\end{figure}

We  will  consider first the case of a non-simply connected observed region which is made of  two caps, $A$, and $B$  centered  at the North and South poles, respectively, see Fig. \ref{fig:sphere-stripe} for their positions on the sphere and their respective stereographic projections. The determinants for the Laplacian are calculated in Appendix \ref{app:sph-caps}, and yield
\begin{equation}
\det \Delta_A \det \Delta_B \det \Delta_C \\
=\big[\det \Delta_{\rm hemisphere} \big]^2  \frac{1}{\pi} \left(\ln \frac{r_2}{r_1}\right) \prod_{n \ge 1} \big[ 1- (\frac{r_1}{r_2})^{2n} \big]^2
\end{equation}
where $r_1$ and $r_2$ are inner and outer radii of the stereographically projected region $C$. The winding function $W(n)$ is conformally invariant, and thus can be evaluated in the annular geometry. Therefore, for this geometry the constant term of the EE becomes
\begin{equation}
S_0= \frac{1}{2}\ln\left(\frac{\det \Delta_A \det \Delta_B \det \Delta_C}{\frac{1}{A}\det \Delta_{A\cup B \cup C} } \right)+ \ln\left(\sqrt{4\pi g} R_c\right) -\frac{1}{2} - W'(1).
\end{equation}

In the  limits of interest, the constant term of the EE, expressed in terms of the  radii $r_1$ and $r_2$ of the stereographic projection of regions $A$ and $B$, becomes
\begin{equation}
S_0( A\cup B ) =  \left\lbrace  
             \begin{aligned}
               &2\Big(\ln \left(\sqrt{8\pi g} R_c\right)  - \frac{1}{2}\Big)- 2 \Big( \frac{r_1}{r_2}\Big)^{\frac{1}{4\pi gR_c^2}}- \Big( \frac{r_1}{r_2}\Big)^{2},   & \quad \textrm{for}\;  r_2 \gg r_1 \\
               &-\frac{\pi}{24} \left(\frac{2\pi r_1}{r_2-r_1}\right)+\ln \left(\sqrt{8\pi g} R_c \right) - \frac{1}{2},   & \quad \textrm{for}\; r_2 \sim r_1 \\
             \end{aligned} \right.  
\end{equation}
In these two limits, in terms of the stereographically projected variables, the obtained values of  $S_0$ are the same as that on the infinite plane. In the limit when two spherical caps are small and are far away from each other, i.e. $r_2\gg r_1$, the scaling function part in $S_0$ vanishes, and we have $S_0=2\Big(\ln \left(\sqrt{8\pi g} R_c\right)  - \frac{1}{2}\Big)$. The constant term suggests that EE in this theory encodes some non-local topological information of the entire system. Since the quantum Lifshitz model is a highly entangled model at the critical point,  it is reasonable that the EE is able to detect the topological information of the surgery. This will be more transparent in the next Section when we consider the case of  plane with multiple small holes.

\subsection{Plane with Multiple Small Holes}
\label{multi_hole}
In this section, we investigate a region on the plane with many small punctured holes. Here, the subsystem $A$ is composed of $m$ small holes. 

\begin{figure}[h]
\centering
\includegraphics[scale = 1.0]{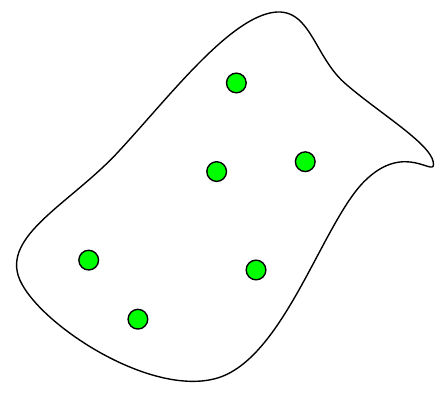}
\caption{Plane with many small holes. The characteristic size of the holes is much smaller than their pairwise distances and of their distance to all boundaries. }
\label{fig:many-hole}
\end{figure}

In general, the $W$ function for the topological sector always depends on the surgery on entire system and it will mix up with the shape/size dependent determinants. Hence it is hard to calculate the  $W$ function for a general case.  However, if we assume that all the disks are small and the distance between them is much larger than their radii, $W(n)$ can be obtained by a Gaussian approximation (details can be found in Appendix \ref{app:W-punc}), leading to the result
\begin{equation}
-W'(1) = m  \Big[ \ln ( 4\pi \sqrt{g} R_c ) - \frac{1}{2} \Big]  + \frac{1}{2} \ln \det G 
\end{equation}
where the elements of the Gram matrix $G$ is for different classical solutions taking unit values on cuts $a$ and $b$
\begin{equation}
G_{ab} = \frac{1}{2\pi}  \int d^2 x  \nabla \phi_{a} \cdot \nabla \phi_b \qquad \phi_a\big|_{{\rm cut }\,\, b} = \delta_{ab}.
\end{equation}

So, in this case, the $S_0$ is found to be 
\begin{equation}
  S_0 =  m  \Big[ \ln ( 4\pi \sqrt{g} R_c ) - \frac{1}{2} \Big] + \frac{1}{2}\big[ \ln \det G +  \ln\frac{\det(A) \det(B)}{\det( A\cup B) } ) \big]
\end{equation}
We have used brackets to separate the two terms in $S_0$. The first term is linear in $\ln R_c$ and the coefficient is proportional to the number of holes, $m$ i.e. the homotopy class of the region $B$.  Its significance lies in the fact that it reflects the topology of the partition of entanglement cut.

Although we cannot calculate the second term in $S_0$ exactly for an arbitrary geometry, we show that it is free from divergences in the thermodynamic limit. To see this, we first write down total EE as
\begin{equation}
S = \alpha \frac{l_{\text{cut}}}{\epsilon} + S_0 .
\end{equation}
In the first term, the area-law term, for a geometry with multiple  holes $l_{\text{cut}}$  is just the sum of their circumferences. For small holes $l_{\text{cut}} \rightarrow 0$. In addition, the degrees of freedom of the smaller subsystem scales as the area $m l_{\text{cut}}^2$. Consequently, the total EE, bounded by logarithm of the total degrees of freedom, will also go to zero. Therefore, in the thermodynamic limit
\begin{equation}
S_0 = S - \frac{l_{\text{cut}}}{\epsilon} \rightarrow \mathcal{O}(1)
\end{equation}
has  a finite value in the limit in which the  holes are small. The parameters for this small hole problem are the shape of the whole boundary and locations of each hole. The spectrum of Laplacian is determined by the shape of the region (``hearing the shape of a drum''), and this information is partially inherited in EE through the determinant term. Various cross ratios parameterizing the locations of the hole will also enter into the EE. This is indeed the case in all the examples we have calculated in previous sections.

\section{Mutual information}
\label{sec:mu}

\subsection{Mutual information on the infinite plane}
\label{sec:mutual-plane}

We will consider first the case of non-simply connected regions of the plane and compute the mutual information for this geometry.
Using the results of entanglement entropy on disk in Section \ref{sec:disk} and Section \ref{sec:annulus}, we can now compute  the mutual information $I(A,B)$ of regions $A$ and $B$ across the annular region $C$ shown in Fig. \ref{fig:ann-conf}. We find
\begin{equation}
\label{eq:I-annul}
I(A,B) = S(A)+S(B) - S(C) = 
\left\lbrace
\begin{aligned}
    2\left(\frac{r_1}{r_2}\right)^{\frac{1}{4\pi gR_c^2}}+\left(\frac{r_1}{r_2}\right)^2,  & \quad \text{thick annulus} \\
   \frac{\pi}{24} \left(\frac{2\pi r_1}{r_2-r_1}\right)+\ln(\sqrt{8\pi g} R_c)- \frac{1}{2}, & \quad \text{thin annulus}   \\
\end{aligned} \right. 
\end{equation}
As expected, the mutual information is a universal scaling function involving explicitly the boson compactification radius $R_c$. Notice that, in the regime in which the annular region $C$ is very thin, it has the same divergent term (but with opposite sign) found before for $S_0$ of the annular region. 

\subsubsection{Thin annulus limit}
\label{sec:thin-annulus-mutual}

Casini {\it et al.}\cite{casini_mutual_2015} proposed a mutual information regulator to extract universal
terms from the entanglement entropy in the continuum limit. They considered the mutual information between regions interior and exterior  to a thin ring of size $\epsilon$ in two dimensions. They argued that, after subtracting a diverging term, in the limit $\epsilon \rightarrow 0$ the remaining term should equal to twice the universal (finite) term in the EE. 
 Since it is not always possible to extract the universal term from entanglement entropy if the UV regulator is, for example, a lattice regulator, the mutual information serves as a useful replacement. In this paper our regulators are perfectly compatible with the geometric regulator requirements outlined in Ref. \cite{casini_mutual_2015}, and so we can use both entanglement and mutual information to extract the universal term of interest. The main focus in Ref. \cite{casini_mutual_2015}  was on the  RG flows in relativistic theories where this universal term serves as an RG monotone.  The theory we study here is non-relativistic (it has $z=2$) nevertheless we expect that this ``mutual information regulator'' should also be useful in this case. 

For a circular region, in order to establish the agreement between the universal terms in the mutual information and in the entanglement entropy,  it turns out the requirement is that 
the constant term in the entanglement entropy for the thin ring, $S({\rm thin\,\,ring})$, is to vanish in the limit  $\epsilon \rightarrow 0$. 
We will see that this does not happen in the model under consideration and,hence, the mutual information is not a suitable regulator for entanglement in this case. More specifically 
we repeated the procedure outlined in Ref. \cite{casini_mutual_2015} for the quantum Lifshitz model  by taking the middle point of the ring C in Fig. \ref{fig:ann-conf} as the boundary of regions $A$ and $B$. Taking the thickness of the ring $\epsilon$ as a physical regulator scale, we remove the $\frac{1}{\epsilon}$ divergent piece in $I(A,B)$ in the thin limit in Eq. \eqref{eq:I-annul}. The mutual information becomes
\begin{equation}
I(A,B) = \ln(\sqrt{8\pi g} R_c)- \frac{1}{2},  \quad \text{thin annulus}
\end{equation}
The constant term is equal to the EE and not \emph{twice} the universal term in the EE of a circle.
This fact can be traced to the fact that the EE of region $C$ has a constant term $\ln \sqrt{8\pi g}R_c - \frac{1}{2}$ even when it is very thin.

Following Refs.  \cite{grover2011entanglement,liu2013refinement,casini_mutual_2015}, one can argue that the UV degrees of freedom (at the length scale $\epsilon$) can only contribute quantities that are local and geometric to the entangling surface - but by a scaling argument in 2D it is not
possible (for suitably geometric regulators) to have a constant term coming from here. Thus, the origin of the discrepancy is not the UV entanglement local to the entangling surface, but it encodes instead some non-local correlations. Perhaps this should not be so surprising since we are working in a critical theory with an abundance of non-local correlations - in spite of the lots of evidence (from AdS/CFT and other calculable models) that this does not happen for relativistic theories \cite{casini_mutual_2015}. 
However this discrepancy does happen for topological theories where the constant term in the mutual information actually vanishes (if the UV scale $\epsilon$ is still larger than the correlation length of the topological theory.) 

\subsubsection{Thick annulus limit}
\label{sec:thick-annulus-mutual}

\begin{figure}[h]
\centering
\includegraphics[scale = 0.5]{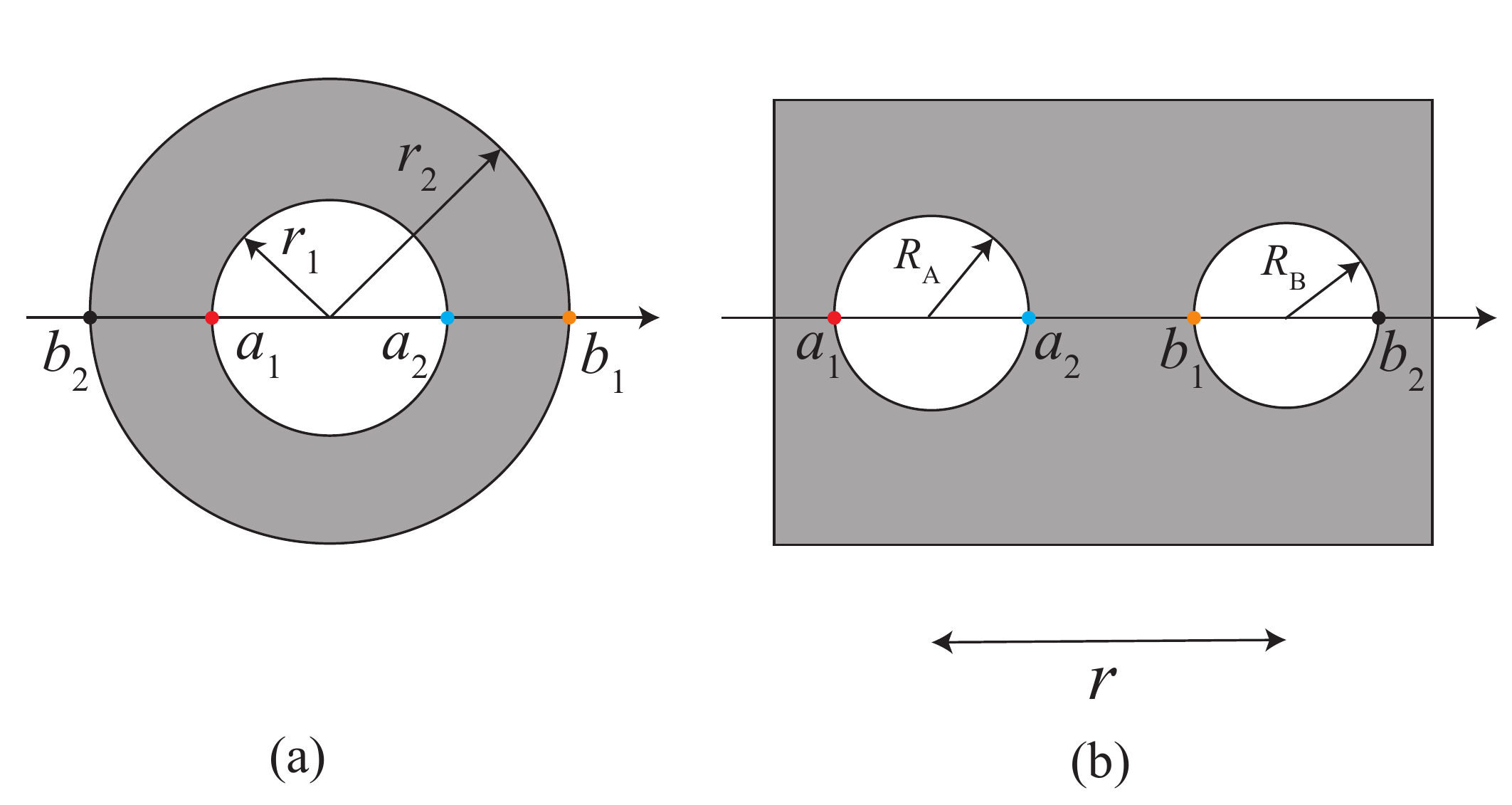}
\caption{(a) Parameter for the annulus configuration. (b) Parameter for two small disjoint circles far away from each other. (a) and (b) are connected through the inversion transformation and the gray region in (a) is mapped to the outside of two circles in (b).}
\label{fig:conf_map}
\end{figure}

On the other hand, when the annular region $C$ is very large, the mutual information simply vanishes as a scaling function of the ratio $R=r_1/r_2$. Notice that the ratio $R$ is a conformally invariant quantity and  it is related to the more familiar cross ratio $x$ in the following way (see Fig.\ref{fig:conf_map}),
\begin{align}
x=\frac{(a_1-b_1)(a_2 - b_2)}{(a_2-b_1)(a_1-b_2)}= \frac{(r_1 + r_2)^2}{(r_2 - r_1)^2} = \frac{(1+R)^2}{(1-R)^2}
\end{align}
The annulus can be mapped to two disjoint circles through a M\"obius transformation. We calculate the cross ratio in Fig.\ref{fig:conf_map} (b),
\begin{equation}
 x=\frac{r^2- (R_A- R_B)^2}{r^2 - (R_A + R_B)^2}
\end{equation}
and equate this to the one obtained in Fig.\ref{fig:conf_map} (a), 
\begin{equation}
\frac{R_A R_B}{r^2 - (R_A+R_B)^2} = \frac{R}{(1-R)^2}.
\end{equation}
In the limit $r_2\gg r_1$, we approximate $R$ as the ratio of various radii,
\begin{equation}
R \simeq \frac{R_A R_B}{r^2}
\end{equation}

Since the mutual information for the thick annulus is a scaling function of the ratio $R$, we expect that it is invariant under conformal mapping. Indeed, according to equation \eqref{eq:I-annul}  for two small distant circles the mutual information scales as\\
\begin{align}
I(A,B) \simeq 2R^{2\Delta_1}+R^{2\Delta_2}
\label{eq:mu_small_circle}
\end{align}
where the exponents $\Delta_1=\frac{1}{8\pi g R_c^2}$ and $\Delta_2=1$.

This result can be compared with that of the relativistic field theories. Cardy \cite{cardy_results_2013} proposed and calculated the expansion of $I(A,B)$ for far disjoint spherical regions of $d+1$ dimensional free scalar field theory. The leading order term is
\begin{equation}
I(A, B) \sim g \left(\frac{R_AR_B}{r^2}\right)^{d-1} ,
\end{equation}
where the separation $r$ is far greater than the radii $R_A$ and $R_B$ of spherical regions. For $d = 2$, Cardy finds $g = \frac{1}{3}$. More generally one expects for any interacting relativistic CFTs:
\begin{equation}
\label{2dmin}
I(A, B) \sim g_\Delta \left(\frac{R_A R_B}{r^2}\right)^{2\Delta_{\rm min}} ,
\end{equation}
where $\Delta_{\rm min}$ is the lowest scaling dimension of a local operator which is not the identity. Indeed this expectation was confirmed in Ref. \cite{Agon:2015ftl} where in addition $g_\Delta$ was computed exactly.

For the quantum Lifshitz model with $z=2$, the mutual information of two small distant circles is shown in Eq.\eqref{eq:mu_small_circle}. Notice that $\Delta_1$ matches the scaling dimension of the vertex operator $V(\vec{x})=\exp(i\phi(\vec{x}))$ and $\Delta_2$ is the scaling dimension of the current operator $\partial_x\phi(\vec{x})$ \cite{Fradkin-2004,Fradkin2013}. $\Delta_{\rm min}$ is determined by the minimal value of $\Delta_1$ and $\Delta_2$ and therefore it depends on the value of the compactification radius $R_c$. These results are analogous to those found in  \cite{headrick2010entanglement,Calabrese:2010he} and indicates that the scaling of mutual information for far disjoint regions is determined by the lowest scaling dimension of primary operator.

We can give an explanation for the result \eqref{2dmin} along the lines of the relativistic CFT
case. The distant small holes can be viewed as a linear superposition of primary fields and the mutual information extracts the correlations between them. As in the relativistic CFTs this follows from an application of the operator product expansion (OPE) to non-local operators (here represented by  higher dimension twist operators) where, viewed from long distances, we can replace the non-local operator with a sum over local ones.   Note the local operators need to be exchanged
in pairs, which gives rise to the fall off of $r^{-4\Delta_{\rm min}}$ and not the naively
expected $r^{-2\Delta_{\rm min}}$ which would be the answer for the R\'enyi entropies
away from $n\neq 1$. In the relativistic case this arises because the squared OPE coefficient  for the non-local operators scales as $(n-1)^2$ for single operator exchanges and thus vanish
after dividing by $(n-1)$ and taking the entanglement entropy limit. Here we expect a similar reasoning. 

\subsection{Mutual information on the sphere}
\label{sec:mutual-sphere}

The above calculation of the mutual information for annulus is done on the infinite plane. Similarly, using the results of EE on the sphere in Section \ref{sec:spherical-cap} and section \ref{sec:two_cap}, we can also compute the mutual information between two spherical caps $A$ and $B$ shown in Fig.\ref{fig:sphere-stripe}, where the two caps are sitting at the North Pole and at the South Pole. 
In terms of the parameters of the stereographically projected region $C$, the mutual information is
\begin{equation}
\label{eq:s-annul}
I(A, B ) = 
\left\lbrace
  \begin{aligned}
    & 2 \Big( \frac{r_1}{r_2}\Big)^{\frac{1}{4\pi gR_c^2}} +\left(\frac{r_1}{r_2}\right)^{2} & \quad r_2 \gg r_1 \\
    & \frac{\pi^2}{12 \ln \frac{r_2}{r_1}}+\ln \left(\sqrt{8\pi g} R_c \right) - \frac{1}{2}  & \quad r_2 \sim r_1 \\
  \end{aligned} \right.  .
\end{equation}
which has the same expression as that for concentrical circles on the infinite plane.

We now discuss the mutual information for two general caps on the sphere. We first fix the cap $B$ to be at the South Pole and let $A$ move away from the North Pole. After a stereographic projection, this geometry maps to an annular region made of two non-concentric circles.
\begin{figure}[hbt]
\includegraphics[scale = 1.0]{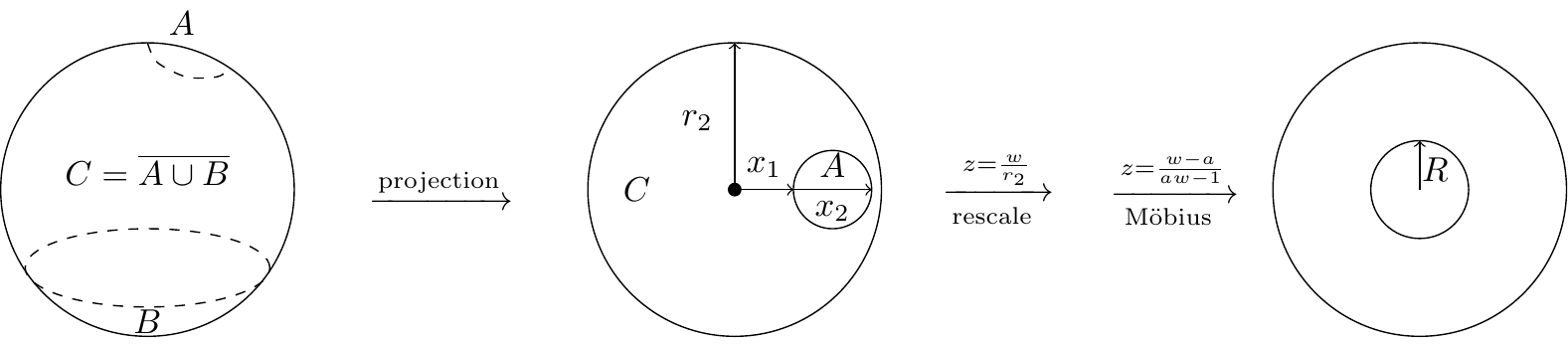}
\caption{Conformal mapping from non-concentric spherical cap to annulus. First do the standard stereographic projection, then rescale the figure by $\frac{1}{r_2}$ such that the outer radius is $1$. Finally, apply a conformal transform $w = \frac{z-a}{az-1}$ that maps the region to an annulus $R < |z| < 1$.}
\label{fig:text-conformal-3-step}
\end{figure}
Fig. \ref{fig:text-conformal-3-step} shows the three steps of conformal mapping that we apply to the non-concentric circles. In the re-scaling operation, the outer radius of the projected circle is set be $1$, hence the only conformal invariant, the cross ratio, is a function of the inner radius $R$
\begin{equation}
  x = \frac{(1+R)^2}{(1-R)^2}
\end{equation}
which becomes $R = \frac{r_1}{r_2}$ in the concentric case. This suggests that the solution should have the same form with a replacement of $\frac{r_1}{r_2}$ by $R$ in the concentric solution. In Appendix \ref{app:sph-caps}, we show that this is indeed the correct procedure by explicit calculation of the determinants. In fact, the regulator-independent mutual information should be (and is) conformally invariant. 

\section{Conclusion and remarks}
\label{sec:summary}

In this paper, we studied the constant (and universal) subleading corrections $S_0$ to the von Neumann entanglement entropy and the mutual information $I(A,B)$ of the quantum Lifshitz model on various geometries by using the replica trick method on the wave function. We obtained the full analytic form of  $S_0$ and $I(A,B)$ and found that both  are conformally invariant and that they include two parts: a cutoff-independent scaling function, and a constant term that depends explicitly (and universally) on the  compactification radius $R_c$.
We reproduced and refined the universal terms in the EE found in Ref. \cite{hsu_universal_2009} for the cylinder and disk geometries. We generalized this method to the cases of caps on the sphere and found that the result is the same as the disk entanglement on the infinite plane. We further studied the case of a subsystem with multiple disjoint holes.  In the small hole limit, we noticed that the coefficient of the universal $ \ln R_c$ term is proportional to the number of holes in the subsystem. This is a demonstration that EE manages to detect non-local information in a critical strongly correlated system.  

Furthermore, we analyzed the behavior of  the mutual information $I(A,B)$ for two distant small circles and found it has power law scaling. The leading order power law exponent in the vanishing mutual information is proportional to the lowest scaling dimension of the primary fields (excluding the identity), which is also the case in relativistic CFTs.  In this case the vertex operator or the current operatorgives rise to the leading power law decay. It would be interesting to generalize this result to other conformal invariant critical point, such as Ising or Potts CFT wave functions

We close with a few comments on several issues.
The entanglement entropy of the quantum Lifshitz model has many interesting parallels with that of compact $U(1)$ Maxwell theory in $2+1$ dimensions  studied in Ref. \cite{agon2014disk} using the dual formulation as a compact boson (see also Refs. \cite{metlitski2011entanglement,radicevic2015entanglement,casini2014entanglement}.) This is not a scale invariant theory so the entanglement of a spherical region actually runs logarithmically
with the size of the observed region.
 However, although this logarithmic term arises due to the compactness of the dual scalar theory and the details of the winding sector sums that produce the logarithm are reminiscent of the winding sector sums found in this paper, the differences between the two theories are significant. In fact the compactified relativistic boson must be regarded as a theory of the IR fixed point of a field theory with a spontaneously broken $U(1)$ symmetry of which the compactified boson is its Goldstone mode. In contrast, the compactified quantum Lifshitz model is physically the quantum phase transition between a uniform phase and a phase with broken translation invariance (for instance, in quantum dimer models \cite{Rokhsar1988,Ardonne-2004}) and is instead a UV fixed point.
 
In fact  the quantum Lifshitz model is also a compact free boson model and it is  dual to a compact gauge field theory in 2+1 dimensions through the relation $\epsilon^{\mu\nu\lambda}\partial_{\nu}a_{\lambda}=\partial^{\mu}\phi$. The wave functional in the ``electric field'' representation can be written as \cite{Fradkin2013}
\begin{equation}
\label{eq:gs_gauge}
|\psi\rangle=\frac{1}{\sqrt{\mathcal{Z}}}\int \mathcal{D}{\bm E} \; e^{-\frac{g}{2}\int d^2x \; {\bm E}^2(\bm x)}\prod_x\delta({\bm \nabla}\cdot{\bm E}(x))|{\bm E}(x)\rangle
\end{equation}
where $\vec{E}=(E_1,E_2)=\partial_0 {\bm a}$ is the ``electric field''.  The factor $\delta(\nabla\cdot\vec{E}(x))$ is the Gauss law constraint (for a system without sources) and requires that the electric field line   form  closed loops on the two dimensional plane. 
Notice that in this representation the wave function looks like a wave function on loop configurations \cite{Freedman-2004}, which is a well known fact in the context of quantum dimer models \cite{Rokhsar1988,Moessner-2001b}.
Since the quantum Lifshitz model is a gauge theory, it might be expected that there is  long range entanglement in the ground state wave function, which is  reminiscent of the topological entanglement entropy (TEE).
Indeed the Gauss' law constraint is crucial to understanding the TEE since otherwise the wave function would be completely local and would not have any long-range entanglement;
 even in gapless theories this
can lead to new and unexpected non-local entanglement 
 \cite{pretko2015}.  Note that it is  an important future task to calculate the mutual information in the compact $2+1$ dimensional $U(1)$  Maxwell theory and related theories to compare to the results of this paper. 

Although the quantum Lifshitz model has dynamical exponents $z=2$, the scaling behavior of EE shows many similarities with the relativistic CFT. It will be interesting to study the effects of local and global quench \cite{calabrese_quantum_2007,cardy_quantum_2015,he_quantum_2014,Nozaki2014} and to do comparisons between this model and the relativistic theory \cite{calabrese_quantum_2016}. We also leave this as a future project.

\begin{acknowledgements}
TZ would like to acknowledge useful discussions with Michael Stone, Israel Klich and Dean Carmi. This work was supported in part by the National Science Foundation grants number NSF-DMR-13-06011 (TZ) and DMR-1408713 (XC,EF) at the University of Illinois. TF is supported by the DARPA, YFA Grant No. D15AP00108.
\end{acknowledgements}

\appendix  
\section{Transformation of $\zeta$ Function Regularized Determinant under a Conformal Mapping}
\label{app:conformal-flow}

Weisberger\cite{weisberger_conformal_1987} developed a method to compute the $\zeta$ function regularized determinant of 2D Laplacian by a Weyl invariant. We summarize his calculation in the case the Dirichlet boundary condition used in this paper. 

Suppose there are two conformally related metrics $\hat{g}_{ab}$ and $e^{2\phi} \hat{g}_{ab}$ on region $\Omega$ of the Riemannian manifold. What is the relation of $\det \Delta$ on region $\Omega$ for this two metrics? One can answer this question by constructing a one parameter family of metrics $g_{ab} = e^{2\phi(t)} \hat{g}_{ab}$ and study the evolution of $\det \Delta$ under this flow. For an infinitesimal Weyl transformation $g_{ab} \rightarrow (1 + 2\delta \phi) g_{ab} $, the spectrum of the Laplacian operator $\Delta = -\frac{1}{\sqrt{g}} \partial_a ( \sqrt{g} g^{ab} \partial_b )$ has the change 
\begin{equation}
\Delta \rightarrow ( 1- 2 \delta \phi ) \Delta , \qquad \delta\lambda_i = \langle \psi_i |\delta \Delta |\psi_i \rangle  = -2 \langle \psi_i |\delta \phi |\psi_i \rangle \lambda _i.  
\end{equation}
In this way,  the change of $\ln \det \Delta$ is expressed as 
\begin{equation}
\begin{aligned}
\delta ( \ln \det \Delta ) &=  \sum_i \frac{\delta \lambda_i}{\lambda_i} = - 2 \lim_{s\rightarrow 0 }   \sum_i \langle  \psi_i |\delta \phi |\psi_i \rangle \lambda_i^{-s} \\
&= -2 \lim_{s\rightarrow 0 }  \frac{1}{\Gamma(s)} \int_0^{\infty} t^{s-1}  \sum_i e^{-\lambda_i t}  \langle  \psi_i |\delta \phi |\psi_i \rangle dt 
\end{aligned}
\end{equation}
The integral can be written as the trace of the heat kernel (the heat kernel representation of the zeta function, see \cite{elizalde_zeta_1994} for example),
\begin{equation}
\int_0^{\infty} \cdots dt = \int_0^{\infty} t^{s-1}  \text{tr}(\delta\phi e^{t \Delta }  )dt  = \mathcal{M}[  \text{tr}(\delta\phi e^{t \Delta } ) ](s)
\end{equation}
where $\mathcal{M}$ represents the Mellin transform. Due to the singularity of $\Gamma(s) = \frac{\Gamma( s+ 1)}{s}$ at $s =0$, the limit actually extracts the coefficient of $\frac{1}{s}$ or, in other words, the residue
\begin{equation}
\delta ( \ln \det \Delta )  = -2\text{res}\big\{ \mathcal{M}[  \text{tr}(\delta\phi e^{t \Delta } ) ](s), s= 0  \big\}.
\end{equation}
The pole structure of the Mellin transform is determined by the small $t$ behavior of the original function (direct mapping theorem). For example, if $f(t) \sim \sum_{n=0}^{\infty} c_n t^n$ when $t \rightarrow 0$, then 
\begin{equation}
\mathcal{M}[f](s) = \int_0^{1}  t^{s-1} [f(t) - \sum_{n=0}^{\infty} c_n t^n]dt +  \int_0^{1} c_n t^{s+n-1} dt +  \int_1^{\infty} f(t) t^{s-1} dt.
\end{equation}
The last term is  convergent if $f$ goes sufficiently fast to $0$ at $\infty$. The singularity in the first term is removed, and so it will converge for all $s$. Hence the second term gives the residue of all the poles
\begin{equation}
\mathcal{M} [f](s)  \sim \sum_{n}\frac{c_n}{s+n}.
\end{equation}
The truncation value of upper limit is arbitrary. If we choose $\epsilon$, then all the arguments are the same, except that the second term will become
\begin{equation}
\int_0^{\epsilon} c_n t^{s+n -1} dt = \frac{c_n}{s+n} \epsilon^{s+n } = \frac{c_n}{s+n} [ 1 + (s+n) \ln \epsilon  \cdots],
\end{equation}
Obviously the residue doesn't change. 

In this case, we need the residue at $s = 0$, or equivalently the constant piece in the small $t$ expansion of the heat Kernel. 
\begin{equation}
  \text{tr}(\delta\phi e^{-t \Delta } ) = \int_{\Omega}  dx G( x,x; t ) \delta \phi \sim \frac{1}{8\pi t} \delta A + \frac{1}{8 \sqrt{\pi t}} \delta P +  \frac{1}{12\pi}\delta L + \frac{1}{8\pi}\delta \Theta  + \mathcal{O}(t^{\frac{1}{2}})
\end{equation}
where $A$ and $P$ stands for area and perimeter. $L$ and $\Theta$ will be defined later. One notice the $\frac{1}{t}$ and $\frac{1}{\sqrt{t}}$ part are divergent for $t$ integration at $s = 0$. But the essence of zeta function regularization is to remove those divergences. 

Consequently we keep only the constant part in the expansion, 
\begin{equation}
  \delta ( \ln \det \Delta ) =-\frac{1}{6\pi} \delta L - \frac{1}{4\pi} \delta \Theta
\end{equation}
where 
\begin{equation}
\label{eq:delta-L}
 \delta  L( \phi, g_{ab} ) = \int_{\Omega}  d^2x \, K \delta \phi + \int_{\partial \Omega} ds \, k_g \delta \phi \qquad \Theta = \int_{\partial \Omega}  ds \, k_g 
\end{equation}
$K$ is the Gaussian curvature, $k_g$ is the geodesic curvature. 

Under a finite conformal transformation, we have (Polyakov-Ray-Singer variation formula)
\begin{equation}
  \ln \frac{\det \Delta_2 }{\det \Delta_1} = -\frac{1}{6\pi}( L_2 - L _1 ) -  \frac{1}{4\pi}(\Theta_2 - \Theta_1 ). 
\end{equation}
where subscript $1$ and $2$ represents the two ends of the conformal flow.
Equivalently
\begin{equation}
I = \det \Delta \exp( \frac{1}{6\pi} L + \frac{1}{4\pi} \Theta ) 
\end{equation}
is a Weyl invariant on the flow.

The integration of $\delta L$ is not explicit given equation \eqref{eq:delta-L}: the complication is the simultaneous evolution of $K$ and $k_g$ along the flow. It turns out the path of evolution is irrelevant if we express all quantities in terms of their values at the end of the flow with metric $\hat{g}_{ab}$,
\begin{equation}
  K =  e^{-2\phi}( \hat{K} + \Delta \phi ) \quad k_g = e^{-2\phi} ( \hat{k}_g - (\vec{n} \cdot \nabla ) \phi ) \quad d^2 x =  e^{2\phi}d\hat{x}^2 \quad ds = e^{\phi} d\hat{s}
\end{equation}
where $\vec{n}$ is the inward normal vector on the boundary curve. Plug in these relations, we get( see reference [2] of \cite{weisberger_conformal_1987} for the last term in $L$)
\begin{equation}
\label{eq:L-int}
\begin{aligned}
\delta L &= \int_{\Omega(\hat{g})}  d^2\hat{x} \, \hat{K} \delta\phi + \int_{\partial \Omega(\hat{g})} ds \, \hat{k}_g \delta \phi +  \int d^2 \hat{x} \,\,\delta\phi \Delta \phi - \int  d\hat{s} \,\delta \phi (\vec{n} \cdot \nabla) \phi \\
&= \int_{\Omega(\hat{g})}  d^2\hat{x} \, \hat{K} \delta\phi + \int_{\partial \Omega(\hat{g})} ds \, \hat{k}_g \delta \phi +  \delta\bigg( \frac{1}{2} \int d^2 \hat{x} \,\, g^{ab} \partial_a \phi \partial_b \phi \bigg)\\
\implies L &= \int_{\Omega(\hat{g})}  d^2\hat{x} \, \hat{K} \phi + \int_{\partial \Omega(\hat{g})} ds \, \hat{k}_g \phi +   \frac{1}{2} \int d^2 \hat{x} \,\, \hat{g}^{ab} \partial_a \phi \partial_b \phi 
\end{aligned}
\end{equation}
For convenience, We will drop all the hats as long as we understand that they are evaluated at the end of the flow. In application, we choose $\hat{g}_{ab} = \delta_{ab}$ and $\phi_1 = 0$. So $L_1 = 0$. $L_2$ can be computed by equation \eqref{eq:L-int}.

\section{$\det \Delta$ for a Single Spherical Cap}
\label{app:sph-cap}

Weisberger\cite{weisberger_conformal_1987} demonstrates his idea by computing $\det \Delta$ on a disk. He first cited $\det \Delta$ on hemisphere(eigenfunctions are spherical harmonics with spectra $\lambda = l ( l+1) / r^2$), then stereographically project the hemisphere to a disk. The metric induced from sphere is conformally connected to the Euclidean metric. So the invariant gives the $\det \Delta$ on a disk(Euclidean metric, Dirichlet boundary condition)
\begin{equation}
  \det \Delta_{\text{Disk}} =r^{- \frac{1}{3}} \exp\big\{ - 2 \zeta'( -1) - \frac{5}{12} -\frac{1}{2} \ln 2\pi  + \frac{1}{3} \ln 2 \big\}. 
\end{equation}
This result is true for all radii. 

Here we compute $\det \Delta$ on a spherical cap. To this end, we take a unit sphere and maps the disk of radius $r$ back to the sphere, the resulting region will be a spherical cap with colatitude $\theta = 2 \arctan r $. For later convenience, express $\det \Delta_{\rm Disk}$ in terms of the determinant on (unit) hemisphere 
\begin{equation}
  \det \Delta_{\rm Disk} = r^{-\frac{1}{3}} \det \Delta_{\text{hemisphere}} \exp( \frac{1}{3}\ln 2 -  \frac{2}{3} ) 
\end{equation}

Now we make use of the Weyl invariant to compute the determinant of a spherical cap. Following Appendix \ref{app:conformal-flow}, we set up a conformal flow between the spherical cap and the disk, and use subscript $1$ and $2$ to denote the initial and final state. The Weyl invariant tells us
\begin{equation}
I = \det \Delta_1 \exp( \frac{1}{6\pi }L_1 + \frac{1}{4\pi} \Theta_1 ) = \det \Delta_2 \exp( \frac{1}{6\pi }L_2 + \frac{1}{4\pi} \Theta_2 ) 
\end{equation}
Since disk is the final stage, $\det \Delta_2 = \det \Delta_{\rm Disk}$. In the end of Appendix \ref{app:conformal-flow}, we explained that $L_2 = 0$.

On the spherical cap, $k_g = \cot \theta $ for small circle, 
\begin{equation}
\Theta_1 = \int_{\partial \Omega} ds \, k_g =  2\pi \cos \theta. 
\end{equation}
We compute $L_1$ by the information on the disk, where $K = 0$, $k_g = \frac{1}{r}$. The induced metric from sphere to $\mathbb{C}\cup \{\infty\}$ is
\begin{equation}
g = \frac{4}{(1 + |z|^2)^2} ( dx\otimes dx + dy \otimes dy ) \implies  \phi = \ln \frac{2}{1 + |z|^2}. 
\end{equation}
Thus 
\begin{equation}
L_1 = \int_{\partial  \Omega(\hat{g}) } ds\, \frac{\phi}{r} + \frac{1}{2} \int_{\Omega(\hat{g}) }  dx dy (\nabla \phi)^2  
= 2\pi \big[ \ln 2 - \frac{r^2}{1+ r^2} \big] 
\end{equation}
On the other hand, $L_2= 0$, 
\begin{equation}
\Theta_2 = \int_{\partial \Omega} ds \, \frac{1}{r} =  2\pi . 
\end{equation}
By the invariant
\begin{equation}
\label{eq:det-cap}
\begin{aligned}
\det \Delta_1  &= \det \Delta_2 \exp\bigg\{ \frac{1}{2}( 1 - \cos \theta ) - \frac{1}{3} \big[ \ln 2 - \frac{r^2}{1+r^2}\big] \bigg\}\\
&= \det \Delta_2 \exp\bigg\{   \frac{2}{3}( 1 - \cos \theta ) - \frac{1}{3}\ln 2 \bigg\} \\
&= r^{-\frac{1}{3}} \det ( \Delta_{\rm hemisphere} ) \exp\bigg\{ - \frac{2}{3}\cos \theta  \bigg\}
\end{aligned}
\end{equation}
This result has been calculated by Dowker (equation (15) of \cite{dowker_functional_1994}, he corrected the transcription error of the spherical cap results in equation (8) of \cite{dowker_further_1995})
\begin{equation}
\frac{1}{2} \ln \det \Delta_1  - \frac{1}{2} \ln \det ( \Delta_{\rm hemisphere}) = - \frac{1}{3}\cos \theta - \frac{1}{6} \ln \tan \frac{\theta}{2} 
\end{equation}

\section{$\det \Delta$ on Cylinder and Annulus}
\label{app:det-cyl-ann}

\begin{figure}[h]
\centering
\includegraphics[scale = 1.0]{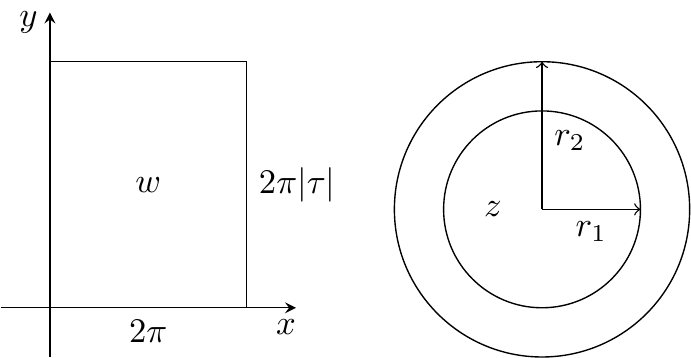}
\caption{Conformal map that transform cylinder to annulus.}
\label{fig:conf-cyl-ann}
\end{figure}

Take a rectangle $[0,2\pi]\times[0,2\pi |\tau|]$ on the $w=(u,v)$ plane and identify $u=0$ and $u = 2\pi$ to make a cylinder. The eigenvalue for Dirichlet boundary conditions on $v = 0, 2\pi |\tau|$ is 
\begin{equation}
 \lambda = m^2  + \frac{n^2}{ |2\tau|^2} , \quad m\in \mathbb{Z}, n\ge 1 
\end{equation}
The spectral zeta function becomes
\begin{equation}
\zeta_{\rm cylinder}( s, \tau) =  \sum_{m\in \mathbb{Z}} \sum_{n > 0}   \frac{1}{|m - \frac{n}{2\tau}|^{2s}} 
= \frac{1}{2}[ \zeta_E( s, -\frac{1}{2\tau} ) - 2 \zeta( 2s ) ].
\end{equation}
where $\zeta_{E}$ is the 2D homogeneous Epstein function
\begin{equation}
\begin{aligned}
  \zeta_E( s, \tau  )  &= \sideset{}{'}\sum_{m,n \in \mathbb{Z}}\frac{1}{|m + n \tau|^{2s}} \\
\end{aligned}
\end{equation}
The analytic continuation of $\zeta_E$ can be found for example in Section 10.2 of the book of Di Francesco {\it et al.} \cite{difrancesco_conformal_2012}. 

Taking a derivative and using modular transform of eta function $\eta(- \frac{1}{\tau}) = \sqrt{-i\tau} \eta( \tau) $, we have the determinant
\begin{equation}
\begin{aligned}
\det\Delta_{\rm cylinder} &= \exp \Big\{-\zeta_{\rm cylinder}'( s = 0, \tau) \Big\} = \exp\Big\{ -\frac{1}{2}\zeta_E'( s = 0, -\frac{1}{2\tau} ) +2 \zeta'( s = 0) \Big\} \\
&= \exp\Big\{ \ln  |\eta(- \frac{1}{2\tau})|^2\Big\} = | \eta( - \frac{1}{2\tau})|^2 = 2|\tau| | \eta(2\tau)|^2 
\end{aligned}
\end{equation}
Then consider the conformal map $z = r_2 \exp( i w )$ that maps the cylinder to an annulus. The outer and inner radii satisfies $r_1 = r_2 \exp( -2\pi |\tau| )  = r_2 q$. The $\Theta$ are both zero for the two geometries. The function $L$ is easier to be calculated by taking $\hat{g}_{ab}$ at the cylinder side. In this case, $K = 0$, $k_g =0 $, the conformal factor is
\begin{equation}
  dz  \otimes d\bar{z} = z \bar{z} dw \otimes d\bar{w} \implies \phi  = \frac{1}{2} \ln z \bar{z} = \ln r_2 - v .
\end{equation}
Hence
\begin{equation}
  L = \frac{1}{2}\int (\nabla \phi)^2 du dv = \frac{1}{2} (2\pi ) (2\pi |\tau|) = 2\pi^2 |\tau|.
\end{equation}
The Weyl invariant
\begin{equation}
I = \det \Delta_{\rm cylinder} = \det \Delta_{\rm annulus}\exp( \frac{\pi}{3} |\tau| ) = \det \Delta_{\rm annulus} q^{-\frac{1}{6}}. 
\end{equation}
yields the determinant of the annulus,
\begin{equation}
 \det \Delta_{\rm annulus} = q^{\frac{1}{6}} 2  \frac{1}{2\pi}\Big(\ln \frac{1}{q}\Big) q^{\frac{1}{6}} \prod_{n > 0} ( 1 - q^{2n} )^2   = \frac{1}{\pi} q^{\frac{1}{3}} \Big(\ln \frac{1}{q}\Big) \prod_{n > 0} ( 1 - q^{2n} )^2 
\end{equation}
where $q = \frac{r_1}{r_2}$. 

\section{Determinants Involved in Two Spherical Caps}
\label{app:sph-caps}

\begin{figure}[h]
\centering
\includegraphics[scale = 1.0]{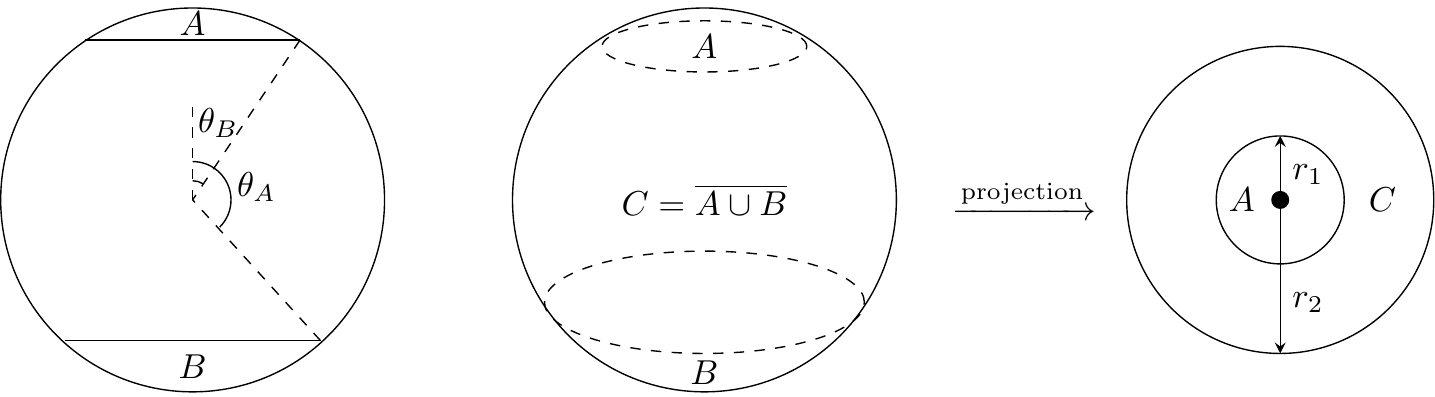}
\caption{Spherical stripe on sphere}
\label{fig:sph-stripe}
\end{figure}

\begin{figure}[h]
\includegraphics[scale = 1.0]{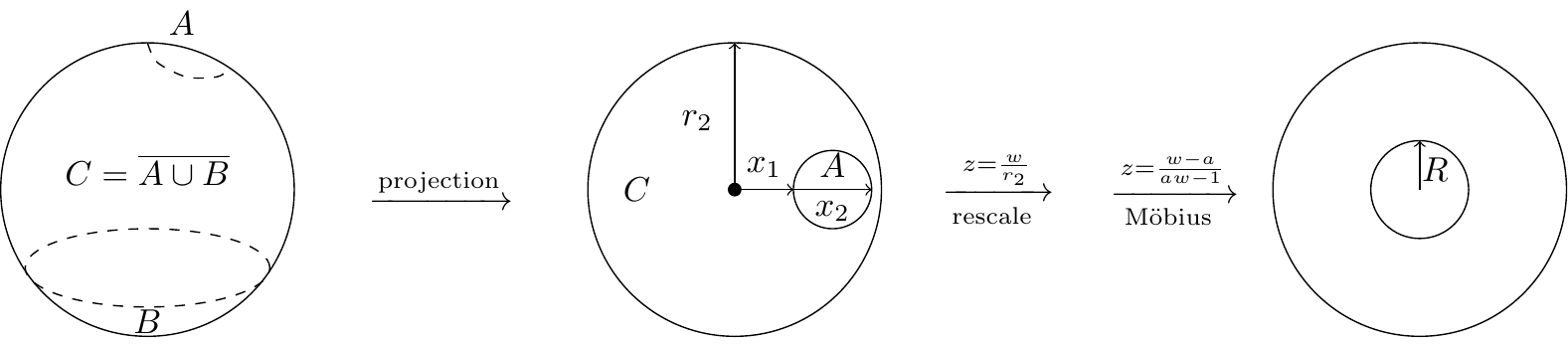}
\caption{Computing determinant of non-concentric configuration. First do the stereographic projection, then rescale the figure by $\frac{1}{r_2}$ such that the outer radius is $1$. Finally, apply a conformal transform $w = \frac{z-a}{az-1}$ that maps the region to an annulus $R < |z| < 1$, where $a = \frac{1 + x_1x_2 - \sqrt{(1-x_1^2 )(1-x_2^2)} }{x_1 + x_2}$, $R = \frac{1 - x_1x_2 - \sqrt{(1-x_1^2 )(1-x_2^2)} }{x_2- x_1} $. In the last step, the variable takes the value after rescaling, i.e. $x_1 = \tan(\frac{\theta_0 - \frac{1}{2}\theta_1}{2})/ \tan \frac{\theta_2}{2}$, $x_2 = \tan(\frac{\theta_0 + \frac{1}{2}\theta_1}{2})/ \tan \frac{\theta_2}{2}$.}
\label{fig:conformal-3-step}
\end{figure}

Consider the region $C$ that is complementary to $A \cup B$ on sphere. A stereographic projection generates a concentric annulus on the plane. The Weyl invariant relates the determinant of $C$ to that of the flat annulus, 
\begin{equation}
\det \Delta( C\text{ on } S^2 ) \exp( \frac{L_1}{6\pi} + \frac{\Delta \Theta}{4\pi}  ) = \det \Delta_{\rm annulus}. 
\end{equation}
The quantity on the exponent can be evaluated as the difference on $C\cup A$ and $A$, 
\begin{equation}
\begin{aligned}
\frac{1}{6\pi}L_1 + \frac{1}{4\pi} \Delta \Theta &= \frac{1}{6\pi}L_1 + \frac{1}{4\pi} \Delta \Theta\bigg|^{C\cup A }_{A}\\
&= \frac{1}{6\pi}2\pi(- \frac{r^2_2}{1+r_2^2} +  \frac{r^2_1}{1+r_1^2}) + \frac{1}{4\pi}2\pi ( \cos\theta_2 - \cos\theta_1 ) \\
& = \frac{2}{3}( \cos\theta_2 - \cos\theta_1).
\end{aligned}
\end{equation}
Hence, 
\begin{equation}
\label{eq:det-annulus}
\det \Delta( C) = \det \Delta_{\rm annulus} \exp \big[ \frac{2}{3}(\cos \theta_1 -\cos \theta_2  ) \big] 
\end{equation}
In particular, the product of the three determinants are 
\begin{equation}
\begin{aligned}
\det \Delta( A) \det \Delta( B) \det \Delta(C ) &= \big[\det ( \Delta_{\rm hemisphere} ) \big]^2 \det \Delta_{\rm annulus} (\frac{r_1}{r_2})^{-\frac{1}{3}}\\
&=\big[\det( \Delta_{\rm hemisphere} ) \big]^2  \frac{1}{\pi} \ln \frac{r_2}{r_1} \prod_{n \ge 1} \big[ 1- (\frac{r_1}{r_2})^{2n} \big]^2 
\end{aligned}
\end{equation}
Now turn to the general case. Suppose region $A$ is tilted at angle $\theta_0$, then the stereographic projection will produce two non-concentric circles for $C\cup A$. In Figure \ref{fig:conformal-3-step}, we show a three-step process to convert it to an annulus.

After doing the stereographic projection, we rescale the radius to $1$. Then apply a M\"obius transformation to center the small circle. We need to specify the parameters in the M\"obius map $z = e^{i \alpha} \frac{w- a}{\bar{a}w - 1}$. Due to the symmetry of the image, we can choose $\alpha = 0$, and a real $a$, such that $(x_1, 0)$ and $(x_2, 0)$ are mapped to $(\pm R, 0)$
\begin{equation}
-\frac{x_1 - a}{ax_1 -1} = \frac{x_2 - a}{ax_2 -1} = R.
\end{equation}
This gives the following solution(for an orientation preserving map $a < 1$ )
\begin{equation}
a = \frac{1 + x_1x_2 - \sqrt{(1-x_1^2 )(1-x_2^2)} }{x_1 + x_2} \qquad R  = \frac{1 - x_1x_2 - \sqrt{(1-x_1^2 )(1-x_2^2)} }{x_2- x_1} 
\end{equation}

Our goal is the compute the accumulated $\exp( \frac{L}{6\pi} + \frac{\Theta}{4\pi})$ factor in these three processes. For convenience, we always take flat metric in the image as the gauge choice. 

First of all, step 3 does not change the product of determinants of $A$ and $C$, so in this section we use $\Delta$ of a quantity with subscript $1,2,3$ to denote change in one of the steps. We first show $\Delta \Theta_3 =0$. The Gaussian curvature satisfies $-\Delta \phi + \hat{K} = e^{2\phi} K(g)$. Since $\hat{K} = 0, \Delta \phi = 0$, $K = 0$. By Gauss-Bonnet theorem, 
\begin{equation}
\int_M K dA + \int_{\partial M} k_g ds = 2\pi \chi(M)
\end{equation}
the $\int k_g ds$ for both sides of the map are equal to the $2\pi$($\chi$ of the disk is 1), hence no change for $\Theta_3$. Furthermore, this M\"obius transformation is the inverse of itself. If we do the conformal flow twice, we should have $\exp( 2L ) = 1$, so $L $ is also $0$.

The rest two steps are generally easy to compute. For completeness, we record the results of none zero change here, 
\begin{equation}
\Delta L_1 = 2\pi ( \ln 2 - \frac{r_2^2}{1+ r_2^2} ) = 2\pi ( \ln 2 - \frac{1- \cos\theta_2}{2} ) \quad \Delta L_2= - 2\pi \ln r_2 \quad \Delta \Theta_1 = 2\pi ( \cos\theta_2 - 1 ),
\end{equation}
and the results of product of the determinants
\begin{equation}
  \begin{aligned}
    \det( A) &\det(B) \det (C) = \det(\text{disk} ) \det(\text{annulus} ) \det(B) \exp( - \frac{1}{6\pi} \Delta L - \frac{1}{4\pi}\Delta \theta )   \\
 &= [\det ( \Delta_{\rm hemisphere} )]^2 \frac{1}{\pi} \ln \frac{1}{R}\prod_{n\ge 1} [ 1- R^{2n}]^2 
  \end{aligned}
\end{equation}
Therefore we can directly use the result of concentric caps by the effective modular parameter $R$.

\section{Topological Sector of Zero Mode in the Path Integral}
\label{app:topo-sum}
The replica trick derived in the text involves a partition function on the $n$-sheet surface with gluing condition-- the $n$ sheets share the same values on the entanglement cut. Specifically, the numerator of
\begin{equation}
  \text{tr}(\rho^n_A) = \frac{\mathcal{Z}_n(\text{gluing condition})}{\mathcal{Z}^n} .
\end{equation}
requires an integration for the $n$ boson fields
\begin{equation}
\mathcal{Z}_n = \int \prod_i [d\phi_i] \exp( - g \int d^2 x \left({\bm \nabla} \phi_i \right)^2)
\end{equation}
subject to the constraint
\begin{equation}
\label{eq:cl_bd_cond}
\phi_1\big|_{\text{cut}} = \phi_2\big|_{\text{cut}} = \cdots \phi_n\big|_{\text{cut}} = \text{cut}( \vec{x} )\quad \text{mod } 2\pi R_c.  
\end{equation}

The general idea stems from Fradkin and Moore\cite{fradkin_entanglement_2006}, where they performed a unitary rotation in the target space. In the simplest $n = 2$ example
\begin{equation}
\begin{bmatrix}
\bar{\phi}_1 \\
\bar{\phi}_2 \\
\end{bmatrix}
=
\begin{bmatrix}
\frac{1}{\sqrt{2}} & \frac{-1}{\sqrt{2}} \\
\frac{1}{\sqrt{2}} & \frac{1}{\sqrt{2}} \\
\end{bmatrix}
\begin{bmatrix}
\phi_1 \\
\phi_2 \\
\end{bmatrix},
\end{equation}
the values on the entanglement cut cancel in first field $\bar{\phi}_1$ and only appear in the second "center of mass" mode $\bar{\phi}_2$. It is then tempting to assume that the center of mass mode is free, and we have the following factorization
\begin{equation}
\mathcal{Z}_{n} = \mathcal{Z}_{\text{Dirichlet} }^{n-1} \mathcal{Z} 
\end{equation}
However, several authors have spotted the subtleties of the compactification radius \cite{oshikawa_boundary_2010,zaletel_logarithmic_2011}: the target space of the $n$-component compact boson is a hypercubic lattice; the nontrivial unitary rotation rotate this lattice such that the rotated fields $\bar{\phi}$ are not standard compact boson any more. We here present a way proposed by Zaletel, Bardarson and Moore\cite{zaletel_logarithmic_2011} to fix this problem within the path integral formulism. 

The trick is to take care of compactification radius in the separated zero modes. For each of the $n$ boson fields, we do decomposition 
\begin{equation}
\phi( \vec{x} ) = \varphi( \vec{x} ) + \phi_{\text{cl}}( \vec{x} ) 
\end{equation}
where the classical mode is responsible to take the value $\text{cut}(\vec{x} )$ on the entanglement cut 
\begin{equation}
\label{eq:cl-bd}
-\nabla^2 \phi_{\text{cl}}( \vec{x} )  = 0, \quad \phi_{\rm cl}( \vec{x} ) \Big|_{\text{cut}} = \phi( \vec{x} ) \Big|_{\text{cut}} \equiv \text{cut}(\vec{x} ) 
\end{equation}
As a result $\varphi( \vec{x} ) \big|_{\text{cut}} = 0$,  the action decomposes as well 
\begin{equation}
  S[\phi] = S[\phi_{\text{cl}}] + S[\varphi] + 2\int_{\partial \Omega}  \varphi \nabla \phi_{\text{cl}} \cdot d \vec{n}  =  S[\phi_{\text{cl}}] + S[\varphi].
\end{equation}

The constrained partition function thus becomes
\begin{equation}
\begin{aligned}
\mathcal{Z}_n &= \prod_{i=1}^n \int_{\varphi_i|_{\text{cut}} = 0 } [d \varphi_i ] \exp( - S[\varphi_i] )  \sum_{\phi^i_\text{cl}} \exp( - \sum_{i=1}^n S[ \phi^i_{\text{cl}} ] ) \\
&= \mathcal{Z}_{\text{Dirichlet} }^{n}  \sum_{\phi^i_\text{cl}} \exp( - \sum_{i=1}^n S[ \phi^i_{\text{cl}} ] ) 
\end{aligned}
\end{equation}
one should interpret $\sum_{\phi^i_\text{cl}}$ as the summation over all the possible solutions of equation set \eqref{eq:cl-bd}. The rotation is now performed for this set of classical modes 
\begin{equation}
\label{eq:rotation}
\begin{bmatrix}
\bar{\phi}_{\text{cl}}^1 \\
\bar{\phi}_{\text{cl}}^2 \\
\bar{\phi}_{\text{cl}}^3 \\
\cdots\\
\bar{\phi}_{\text{cl}}^{n-1} \\
\bar{\phi}_{\text{cl}}^{n}\\
\end{bmatrix}
=
\begin{bmatrix}
\frac{1}{\sqrt{2}} & \frac{-1}{\sqrt{2}} & & & &\\
\frac{1}{\sqrt{6}} & \frac{1}{\sqrt{6}} & \frac{-2}{\sqrt{6}} & & &\\
\frac{1}{\sqrt{12}} & \frac{1}{\sqrt{12}} & \frac{1}{\sqrt{12}} &  \frac{-3}{\sqrt{12}}& &\\
\cdots & \cdots & \cdots & \cdots & \cdots & \\
\frac{1}{\sqrt{n^2 -n}} & \frac{1}{\sqrt{n^2 -n}} & \cdots & \cdots  & \frac{-(n-1)}{\sqrt{n^2 -n}}&\\
\frac{1}{\sqrt{n}}& \frac{1}{\sqrt{n}}& \frac{1}{\sqrt{n}}& \cdots& \frac{1}{\sqrt{n}}& \\
\end{bmatrix}
\begin{bmatrix}
\phi_{\text{cl}}^1 \\
\phi_{\text{cl}}^2 \\
\phi_{\text{cl}}^3 \\
\cdots\\
\phi_{\text{cl}}^{n-1} \\
\phi_{\text{cl}}^{n}\\
\end{bmatrix}
\end{equation}
which we abbreviate using vector notation $\bar{\vec{\phi}}_\text{cl} = U_n \vec{\phi}_\text{cl} $. The unitary rotation matrix $U_n$ is carefully chosen such that the first $n -1$ rotated classical modes vanishes up to compactification radius on the entanglement cut. 

To write down precisely the boundary condition on the cut, we switch to the case of a single cut, such that the constraint can be associated with a winding vector $\vec{w} \in \mathbb{Z}^{n-1}$
\begin{equation}
\phi^i_\text{cl} (\vec{x})\Big|_\text{cut} = \text{cut}( \vec{x} ) + 2\pi w_i R_c \qquad i = 1, \cdots, n-1, \qquad \phi^n_\text{cl}( \vec{x} ) \Big|_\text{cut} = \text{cut}( \vec{x} ) 
\end{equation}
it is easy to generalize the $\vec{w}$ vector to the general $m$-cut case as in appendix \ref{app:W-punc}. We define the minor matrix resulting from deleting the $n$-th row and column of $U_n$ to be $M_{n-1}$ (it is a $n-1 \times n -1$ matrix), which can also be written as
\begin{equation}
M_{n-1} = \text{diag}( 1, 1, \cdots, \frac{1}{\sqrt{n}} ) U_{n-1} 
\end{equation}
the new boundary condition for the rotated fields is then
\begin{equation}
\begin{aligned}
  \bar{\phi}^i_\text{cl}\Big|_\text{cut} &= 2\pi R_c \big(M_{n-1})_{ij} w_j \qquad \text{for }  \quad i< n  \quad \\
  \bar{\phi}^n_\text{cl}\Big|_\text{cut} &= \sqrt{n} \text{cut}(\vec{x}) + \frac{2\pi R_c}{\sqrt{n}}\sum_{i=1}^{n-1}  w_i .
\end{aligned}
\end{equation}
The $n$-th rotated fields is identified as the center of mass mode since it is defined as $\sum_{i=1}^n\phi_{\text{cl}}^i/\sqrt{n}$.

We can decouple the integration in the zero mode into the summation over the winding vector $\vec{w}$ and cut function $\text{cut}(\vec{x})$
\begin{equation}
\sum_{\phi^i_\text{cl}} \exp( - \sum_{i=1}^n S[ \phi^i_{\text{cl}} ] )  = \sum_{\vec{w}} \exp( - \sum_{i=1}^{n-1} S[ \bar{\phi}^i_\text{cl}] ) \int [d\text{cut}] \exp( - S[ \bar{\phi}^n_\text{cl}] ) 
\end{equation}
the summation over the winding vector is defined to $W$ function 
\begin{equation}
W( n) = \sum_{\vec{w}} \exp( - \sum_{i=1}^{n-1} S[ \bar{\phi}^i_\text{cl}] ) 
\end{equation}
For the other part, we make a change of path integral measure
\begin{equation}
[d\text{cut}(x)]  = [d \frac{1}{\sqrt{n}}(\bar{\phi}_\text{cl}\Big|_\text{cut} - \frac{2\pi R_c}{\sqrt{n}}\sum_{i} w_i)  ] = n^{- \frac{1}{2} \frac{L}{a} } [d  \bar{\phi}_\text{cl}\Big|_\text{cut}].
\end{equation}
where $a$ is a short distance cutoff and $L$ is the length of the cut. The $n^{-\frac{1}{2} \frac{L}{a} }$ factor will only contribute an area law term in the EE; we thus neglect it for the sake of constant term. Finally, we combine the integration of one Dirichlet field and $\bar{\phi}^n_\text{cl}$ to be a free field 
\begin{equation}
\bar{\phi}_n = \varphi_n + \bar{\phi}^n_\text{cl} 
\end{equation}
and claim the integration of the individual parts form a free partition function 
\begin{equation}
\begin{aligned}
\mathcal{Z}_{\text{Dirichlet} } \int [d\bar{\phi}_\text{cl}\big|_\text{cut}]   \exp( - S[ \bar{\phi}^n_\text{cl}] ) 
 &=  \int [d\bar{\phi}_\text{cl}\big|_\text{cut}]  [d\varphi_n ]   \exp( - S[ \bar{\phi}^n_\text{cl}] - S[ \varphi_n ] ) \\
 &= \int [d\bar{\phi}_n ]   \exp( - S[ \bar{\phi}_n])   ) \\
 &= \mathcal{Z}_{\rm Free}
\end{aligned}
\end{equation}

This decomposition is easily generalized to more complicated entanglement cut structures. We therefore obtain the general formula
\begin{equation}
\text{tr}(\rho_A^n) = \Big[\frac{ \mathcal{Z}_{\text{Dirichlet} }  }{\mathcal{Z}}\Big]^{n-1} W(n ) 
\end{equation}

\section{$W$ for Semi-Infinite Cylinder}
\label{app:W-cyld}

In this appendix, we compute the $W$ function of cylinder geometry\cite{zaletel_logarithmic_2011}. Compared with \cite{zaletel_logarithmic_2011}, here we give the full analytically continued expression $W'(1)$, which depends on all the operator contents of the ground state CFT.

The geometry is a cylinder $[-\pi, \pi]\times[-\pi |\tau|, \pi |\tau|]$ with Dirichlet boundary condition at $ x = \pm \pi |\tau|$ and periodic in $y$. From our analysis in Appendix \ref{app:topo-sum}, we need to find solutions of Laplace equation with boundary conditions
\begin{equation}
\vec{\phi}_{\text{cl}} \big|_{x = 0} = 2\pi R_c M_{n-1} \vec{w} 
\end{equation}
which are
\begin{equation}
\vec{\phi}_{\text{cl}} = 2\pi R_c M_{n-1} \vec{w}  \frac{( \pi |\tau| - |x| )}{\pi |\tau| }
\end{equation}
Therefore the $W$ function is
\begin{equation}
\begin{aligned}
W(n) &=  \sum_{\vec{w}\in \mathbb{Z}^{n-1}}\exp( - g \int d^2 x \nabla \vec{\phi}_{\text{cl}} \cdot \nabla \vec{\phi}_{\text{cl}}  ) \\
&=  \sum_{\vec{w}\in \mathbb{Z}^{n-1}}\exp\big[ - g(\frac{1}{\pi |\tau|})^2 4 \pi^2 |\tau|  4\pi^2 R_c^2 \vec{w}^T M^T_{n-1} M_{n-1} \vec{w}  \big]\\
&= \sum_{\vec{w}\in \mathbb{Z}^{n-1}}\exp\big[ - \pi \frac{16\pi g R_c^2}{|\tau| }\vec{w}^T M^T_{n-1} M_{n-1} \vec{w}  \big]\\
\end{aligned}
\end{equation}

We define the matrix on the exponent to be $T_{n-1}$
\begin{equation}
\begin{aligned}
  T_{n-1} &= M_{n-1}^T M_{n-1} = U^T_{n-1} \diag(1, 1, \cdots, \cdots, \frac{1}{n} ) U_{n-1} \\
  &= 
  \begin{bmatrix}
    1 - \frac{1}{n} & - \frac{1}{n} & - \frac{1}{n } & \cdots & -\frac{1}{n} & \\
    -\frac{1}{n}  & 1- \frac{1}{n} & - \frac{1}{n} & \cdots & -\frac{1}{n} & \\
    \cdots &     \cdots &     \cdots &     \cdots &     \cdots &    \\
    -\frac{1}{n}  & - \frac{1}{n} & \cdots  & 1 - \frac{1}{n} & -\frac{1}{n} & \\
    -\frac{1}{n}  & - \frac{1}{n} & \cdots  & - \frac{1}{n} & 1-\frac{1}{n} & \\
  \end{bmatrix}
\end{aligned}
\end{equation}

We perform the analytic continuation by first applying the reciprocal formula for multidimensional theta function\cite{bellman_reciprocity_1961}
\begin{equation}
\theta( \vec{z}, T ) = \sum_{n} \exp\big\{ - \pi \vec{n}\cdot  T \vec{n} + 2\pi i \vec{n} \cdot  \vec{z}\big\}
 = (\det T)^{-\frac{1}{2}}  \sum_{n} \exp\big\{ - \pi (\vec{n} + \vec{z} ) \cdot  T^{-1} ( \vec{n} + \vec{z} ) \big\}.
\end{equation}
Let $c = \frac{16\pi g R_c^2}{|\tau|}$, then
\begin{equation}
\begin{aligned}
  W(n) &= \sum_{ \vec{w} \in \mathbb{Z}^{n-1} } \exp( -\pi  c \vec{w}^T \cdot T_{n-1} \vec{w} ) = (\det cT_{n-1} )^{- \frac{1}{2} } \sum_{ \vec{w} \in \mathbb{Z}^{n-1} }\exp ( - \pi \frac{1}{c} \vec{w}^T  \cdot T^{-1} _{n-1} \vec{w} ) \\
  &= \sqrt{n } c^{- \frac{n-1}{2} } \sum_{ \vec{w} \in \mathbb{Z}^{n-1} }\exp ( - \pi \frac{1}{c} \vec{w}^T \cdot T^{-1} _{n-1} \vec{w} )
\end{aligned}
\end{equation}
The inverse matrix is independent of $n$ 
\begin{equation}
T_{n-1}^{-1} = 
  \begin{bmatrix}
    2 & 1 & 1 & \cdots & 1 & \\
    1  & 2 & 1 & \cdots & 1 & \\
    \cdots &     \cdots &     \cdots &     \cdots &     \cdots &    \\
    1  & 1 & \cdots  & 2 & 1 & \\
    1  & 1 & \cdots  & 1 & 2 & \\
  \end{bmatrix}
\end{equation}
and so analytic continuation is performed on the \emph{dimension} of the matrix only.

The trick is to complete the square for the exponent
\begin{equation}
\vec{w} \cdot T_{n-1}^{-1} \vec{w} = \sum_{i=1}^{n-1} w_i^2  +  ( \sum_{i=1}^{n-1}  w_i  )^2 
\end{equation}
such that the "center of mass" square can be converted to a delta function
\begin{equation}
\begin{aligned}
\sum_{ \vec{w} \in \mathbb{Z}^{n-1} } & \exp ( - \pi \frac{1}{c} \vec{w}^T \cdot T^{-1} _{n-1} \vec{w} ) = \sum_{ \vec{w} \in \mathbb{Z}^{n-1} } \exp[ -\frac{\pi}{c} \sum_{i=1}^{n-1} w_i^2 - \frac{\pi}{c} (\sum_{i=1}^{n-1} w_i)^2 ] \\
&= \int_{-\infty}^{\infty} dx \sum_{ \vec{w} \in \mathbb{Z}^{n-1} } \exp[ -\frac{\pi}{c} \sum_{i=1}^{n-1} w_i^2 - \frac{\pi}{c} x^2 ] \delta( x - \sum_{i=1}^{n-1} w_i ) \\
&= \int_{-\infty}^{\infty} \frac{dk}{2\pi} \int_{-\infty}^{\infty} dx \sum_{ \vec{w} \in \mathbb{Z}^{n-1} } \exp[ - \frac{\pi}{c} \sum_{i=1}^{n-1} w_i^2 - \frac{\pi}{c} x^2 ] \exp[ ik( x - \sum_{i=1}^{n-1} w_i )  ]\\
&= \int_{-\infty}^{\infty} \frac{dk}{2\pi} \int_{-\infty}^{\infty} dx \big[\sum_{w\in\mathbb{Z}}  \exp( - \frac{\pi}{c}  w^2 -ik w )\big]^{n-1} \exp[ - \frac{\pi}{c} x^2 ] e^{ikx}\\
&= \int_{-\infty}^{\infty} \frac{dk}{2\pi}  \big[\sum_{w\in\mathbb{Z}}  \exp( - \frac{\pi}{c}  w^2 -ik w )\big]^{n-1} \sqrt{c} \exp[ -\frac{c}{4\pi} k^2 ] \\
&= \int_{-\infty}^{\infty} \frac{dk}{\sqrt{\pi}} e^{-k^2}\big[\sum_{w\in\mathbb{Z}}  \exp( - \frac{\pi}{c}  w^2 -2i \sqrt{\frac{\pi}{c}}k w )\big]^{n-1}
\end{aligned}
\end{equation}

We therefore obtain an integral expression for the  $W$ function
\begin{equation}
\label{eq:W-analy-cont}
W(n) = \sqrt{n } c^{ -\frac{n-1}{2} } \int_{-\infty}^{\infty} \frac{dk}{\sqrt{\pi}} e^{-k^2}\big[\sum_{w\in\mathbb{Z}}  \exp( - \frac{\pi}{c} w^2 -2i \sqrt{\frac{\pi}{c}}k w )\big]^{n-1}
\end{equation}
notice that this gives the correct normalization $W(1) = 1$.

Terms that appear in the EE is
\begin{equation}
\label{eq:W-prime-analy-cont}
  -W'(1) =  \ln \sqrt{c}  - \frac{1}{2} - \int_{-\infty}^{\infty} \frac{dk}{\sqrt{\pi}} e^{-k^2} \ln \big[\sum_{w\in\mathbb{Z}}  \exp( - \frac{\pi}{c}  w^2 -2i \sqrt{\frac{\pi}{c}}k w )\big]
\end{equation}

The semi-infinite limit corresponds to $|\tau|\gg 1$ and $c \ll 1$
\begin{equation}
\begin{aligned}
- W'(1) &\simeq \ln \sqrt{c} - \frac{1}{2} - 2\int_{-\infty}^{\infty} \frac{dk}{\sqrt{\pi}} e^{-k^2} \exp( - \frac{\pi}{c} )\cos( 2\sqrt{\frac{\pi}{c}} k ) \\
&=  \ln \sqrt{c} - \frac{1}{2} - 2 \exp(- \frac{2\pi }{c} )\\
&=  - \frac{1}{2} \ln |\tau | +  \ln \sqrt{ 16 \pi g }R_c - \frac{1}{2} - 2 \exp(- \frac{|\tau|}{8 gR_c^2 })
\end{aligned}
\end{equation}
The first three terms can also be obtained from the Gaussian integral approximation of the multi-dimensional theta function. The last term represents the contribution from the primary fields of the ground state CFT.

Notice that quadratic action 
\begin{equation}
S[\phi_{\text{cl}}] = g\int \sqrt{\det(g_{ab} ) } \,  d^2x\,   g^{ab} \partial_a \phi_{\text{cl}} \partial_b \phi_{\text{cl}}
\end{equation}
is invariant under coordinate transformation. Furthermore, a conformal transformation preserve the Laplace equation satisfied by $\phi_{\text{cl}}$. Therefore the $W$ function thus defined is a conformal invariant. In other words, we can compute it in any convenient geometry as long as it is connected to the original one via conformal transformation. 

\section{$W$ on Annulus}
\label{app:W-ann}

Consider the sum over topological sector $W( n)$ for compactified boson on an annulus with Dirichlet boundary conditions on the larger circle. 

The classical solution is proportional to $\ln r$, 
\begin{equation}
\label{eq:ann_sol}
\phi_{\text{cl}} = \frac{\phi_{\text{cl}}( r = r_1 ) }{\ln \frac{r_1}{r_2} } \ln \frac{r}{r_2},
\end{equation}
and hence $W$ function is
\begin{equation}
\begin{aligned}
W(n) &= \sum_{\vec{w}} \exp\bigg\{- g  (2\pi R_c)^2 \vec{w}^T M^T_{n-1} M_{n-1} \vec{w} \int_{r_1}^{r_2}  2\pi r \big(\frac{1}{r\ln \frac{r_1}{r_2}}\big)^2 dr  \bigg\} \\
&= \sum_{\vec{w}} \exp\bigg\{- 2\pi g \frac{(2\pi R_c)^2}{ \ln \frac{r_2}{r_1} }    \vec{w}^T T_{n-1} \vec{w}\bigg\} \\
&= \sum_{\vec{w}} \exp\bigg\{- \pi  \frac{8 \pi^2 g R_c^2 }{ \ln \frac{r_2}{r_1} }    \vec{w}^T T_{n-1} \vec{w}\bigg\} \\
\end{aligned}
\end{equation}

The analytic continuation is performed similarly as in \ref{app:W-cyld}. We quote the general expression in equation \eqref{eq:W-analy-cont}
\begin{equation}
W(n) = \sqrt{n } c^{ -\frac{n-1}{2} } \int_{-\infty}^{\infty} \frac{dk}{\sqrt{\pi}} e^{-k^2}\big[\sum_{w\in\mathbb{Z}}  \exp( - \frac{\pi}{c} w^2 -2i \sqrt{\frac{\pi}{c}}k w )\big]^{n-1}
\end{equation}
where $c$ now takes the value $\frac{8\pi^2 g R_c^2}{\ln \frac{r_2}{r_1}}$ and also equation \eqref{eq:W-prime-analy-cont}
\begin{equation}
\label{eq:W-prime}
  -W'(1) =  \ln \sqrt{c}  - \frac{1}{2} - \int_{-\infty}^{\infty} \frac{dk}{\sqrt{\pi}} e^{-k^2} \ln \big[\sum_{w\in\mathbb{Z}}  \exp( - \frac{\pi}{c}  w^2 -2i \sqrt{\frac{\pi}{c}}k w )\big]
\end{equation}

If $r_2 \gg r_1 $, 
\begin{equation}
\begin{aligned}
-W'(1) &\simeq \ln \sqrt{c} - \frac{1}{2} - 2\int_{-\infty}^{\infty} \frac{dk}{\sqrt{\pi}} e^{-k^2} \exp( - \frac{\pi}{c} )\cos( 2\sqrt{\frac{\pi}{c}} k ) \\
&= \ln \sqrt{c} - \frac{1}{2} - 2 \exp( - \frac{2\pi }{c} ) \\
&= - \frac{1}{2} \ln \ln \frac{r_2}{r_1}  + \ln \sqrt{8\pi^2 g} R_c - \frac{1}{2} - 2\Big( \frac{r_1}{r_2}\Big)^{\frac{1}{4\pi g R_c^2}}
\end{aligned}
\end{equation}

On the other hand, if $r_2 \sim r_1$, we transform equation \eqref{eq:W-prime} by reciprocal formula 
\begin{equation}
\sum_{w\in\mathbb{Z}}  \exp( - \frac{\pi}{c}  w^2 -2i \sqrt{\frac{\pi}{c}}k w ) = \sqrt{c} e^{-k^2} \sum_{ w \in \mathbb{Z}} \exp(- \pi c w^2 + 2 \sqrt{\pi c} w k )
\end{equation}
and hence obtain
\begin{equation}
\begin{aligned}
-W(1)' &= -\int_{-\infty}^{\infty} \frac{dk}{\sqrt{\pi}} e^{-k^2} \ln [ \sum_{ w \in \mathbb{Z}} \exp(- \pi c w^2 + 2 \sqrt{\pi c} w k ) ] \\
& = -\int_{-\infty}^{\infty} \frac{dk}{\sqrt{\pi}} e^{-k^2} \ln [ \sum_{ w \in \mathbb{Z}} \exp(- (\sqrt{\pi c}w - k )^2 + k^2 ) ] \\
& \simeq 0 \qquad \text{when } c \rightarrow \infty
\end{aligned}
\end{equation}

\section{Replica Formula on Sphere}
\label{app:rep_formula_sph}

In this section, we will derive a slightly different replica formula on sphere. 

We start with a single entanglement cut. The derivation follows the same procedure until the quantum-classical decomposition. We rewrite the boundary conditions of classical fields in equation \eqref{eq:cl_bd_cond} as
\begin{equation}
\label{eq:cl_bd_cond_sph}
\phi^j_{\rm cl} \big|_{\text{cut}} = w_j 2 \pi R_c + \text{cut}(\vec{x}) \qquad \text{ for } j < n \qquad \phi^n_{\rm cl}\big|_{\text{cut}} \equiv \text{cut}( \vec{x}  )
\end{equation}
the winding number only gives an overall shift to the classical mode, and hence
\begin{equation}
S[\phi^j_{\rm cl}] = S[\phi^n_{\rm cl} ] = S[ \text{cut}( \vec{x} ) ] \qquad \forall j 
\end{equation}
i.e. no rotation is needed to accommodate the winding number. 

Hence the summation over all the classical fields boils down to a single field
\begin{equation}
\sum_{\phi^i_\text{cl}} \exp( - \sum_{i=1}^n S[ \phi^i_{\text{cl}} ] ) = \sum_{\phi^n_{\text{cl}}} \exp( - n S[\phi_{\text{cl}}^n ] ) 
= \sum_{\phi^n_{\text{cl}}} \exp( - S[\sqrt{n} \phi_{\text{cl}}^n ] ) 
\end{equation}

In Appendix \ref{app:topo-sum}, we claim with Dirichlet boundary condition on the physical edge
\begin{equation}
\mathcal{Z}_{\text{Free}} = \mathcal{Z}_{\text{Dirichlet}} \sum_{\phi^n_{\text{cl}}} \exp( - S[ \bar{\phi}_{\text{cl}}^n ] ) 
= \int [d\varphi_n] \exp( - S[\varphi_n]) \sum_{\phi^n_{\text{cl}}} \exp( - S[\bar{\phi}_{\text{cl}}^n ] ) 
\end{equation}
by identifying the free partition function as integrating over the new field
\begin{equation}
\bar{\phi}_n = \varphi_n + \bar{\phi}_{\text{cl}}^n.
\end{equation}
Heuristically, the $\bar{\phi}_{\text{cl}}^n$ enumerates all possible values on the entanglement cut and $\varphi_n$ exhausts the rest degree of freedom and therefore the combined field has no constraint in the total area. 

By contrast in the spherical case, we have
\begin{equation}
\mathcal{Z}_{\text{Dirichlet}} \sum_{\phi^n_{\text{cl}}} \exp( - S[\sqrt{n} \phi_{\text{cl}}^n ] ) 
= \int [d\varphi_n] \exp( - S[\varphi_n]) \sum_{\phi^n_{\text{cl}}} \exp( - S[\sqrt{n} \phi_{\text{cl}}^n ] ) 
\end{equation}
and the analogous new field is
\begin{equation}
\bar{\phi}_n = \varphi_n + \sqrt{n} \phi^n_{\text{cl}}.
\end{equation}
The difference lies in the fact that the new field $\bar{\phi}_n$ has $2\pi \sqrt{n} R_c$ as its compactification radius. To see this, we shift the nth copy of field $\phi_n $ by $2\pi R_c$ then in the quantum-classical decomposition, only classical part is shifted,
\begin{equation}
\phi_n \rightarrow \phi_n + 2\pi R_c = \varphi_n + ( \phi_{\text{cl}}^n  + 2\pi R_c )
\end{equation}
then the resulting change in $\bar{\phi}_n$ is
\begin{equation}
\bar{\phi}_n \rightarrow \bar{\phi}_n + \sqrt{n} 2\pi R_c 
\end{equation}
Physically, the partition function should be invariant about the shift, thus the compactification radius is amplified by a factor of $\sqrt{n}$, which is summarized in the following equation
\begin{equation}
\mathcal{Z}_{\text{Dirichlet}} \sum_{\phi^n_{\text{cl}}} \exp( - S[\sqrt{n} \phi_{\text{cl}}^n ] )  = \sqrt{n} \mathcal{Z}_{\text{Free}}
\end{equation}
Note that for a region with Dirichlet boundary condition, we are unable to do the global shift!

When we have two entanglement cuts on the sphere, equation \eqref{eq:cl_bd_cond_sph} should be modified as
\begin{equation}
\begin{aligned}
\phi^j_{\rm cl} \big|_{\text{cut 1}} &= w^1_j 2 \pi R_c + \text{cut}(\vec{x}) \qquad \text{ for } j < n \qquad \phi^n_{\rm cl}\big|_{\text{cut 1}} \equiv \text{cut}_1( \vec{x}  ) \\
\phi^j_{\rm cl} \big|_{\text{cut 2}} &= w^2_j 2 \pi R_c + \text{cut}(\vec{x}) \qquad \text{ for } j < n \qquad \phi^n_{\rm cl}\big|_{\text{cut 2}} \equiv \text{cut}_2( \vec{x}  ) .
\end{aligned}
\end{equation}
Again $S[\phi^j_{\rm cl}] = S[\phi^j_{\rm cl} - w^1_j 2\pi R_c]$, we can get rid of one sets of the winding numbers on cut 1; effetively, we can change the boundary condition to
\begin{equation}
\begin{aligned}
\phi^j_{\rm cl} \big|_{\text{cut 1}} &= \text{cut}_1( \vec{x}  )  \qquad \forall j\\
\phi^j_{\rm cl} \big|_{\text{cut 2}} &= w_j 2 \pi R_c + \text{cut}(\vec{x}) \qquad \text{ for } j < n \qquad \phi^n_{\rm cl}\big|_{\text{cut 2}} \equiv \text{cut}_2( \vec{x}  ) 
\end{aligned}
\end{equation}
where $w_j = w^2_j - w^1_j$. Then we perform the rotation as in \eqref{eq:rotation} to separate out the contribution from the $\vec{w}$ vector as $W$ function. The remaining center of mass field $\bar{\phi}_n$ is equal to $\sqrt{n} \text{cut}_{1,2}( \vec{x} )$ on the two entanglement cuts. Hence in combining a Dirichlet partition function with the center of mass field, the compactification radius is again amplified to $\sqrt{n} 2\pi R_c$. In summary, 
\begin{equation}
\mathcal{Z}_n = \sqrt{n}  \mathcal{Z}_{\rm Dirichlet}^{n-1} \mathcal{Z}_{\text{Free}} 
\end{equation}

\section{$W$ on the Punctured Plane}
\label{app:W-punc}

\begin{figure}[h]
\centering
\includegraphics[scale = 1.0]{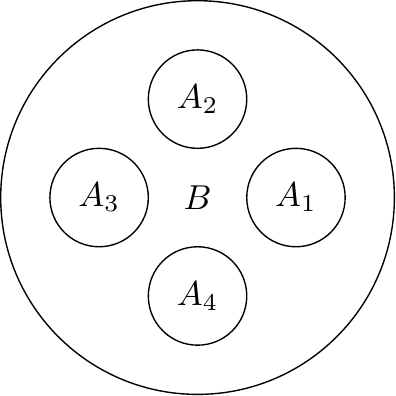}
\label{fig:punc-plane}
\end{figure}
We consider the another planar configuration in which the subsystem $A$ consists disjoint parts $A_a$. Each classical mode corresponds to boundary value problem
\begin{equation}
\nabla^2 \phi = 0\qquad \phi\Big|_{\partial A_a } = 2\pi R_c w^{a} \quad a = 1, 2, \cdots,  m 
\end{equation}
There is a set of basis solutions which takes unit value on each boundary
\begin{equation}
\nabla^2 \phi_a = 0 \quad \phi^{a}\big|_{\partial A_b } = \delta_{ab} \quad a,b = 1, 2, \cdots, m
\end{equation}
and any solution can be expanded as a linear superposition
\begin{equation}
\phi = 2\pi R_c  w^{a} \phi_{a} 
\end{equation}
We obtain the classical field in the replica formula for $W$
\begin{equation}
\phi_{i,\text{cl}} = (M_{n-1})_{ij} (2\pi R_c) w_j^a \phi_a
\end{equation}
where $2\pi R_c w_j^a$ is the boundary value on $\partial A_a $ for the $j$th classical solution. 

For the sake of clarity, we define an inner product for the solutions
\begin{equation}
( \phi_1, \phi_2 ) = \frac{1}{2\pi} \int d^2x \nabla \phi_1 \cdot \nabla \phi_2 
\end{equation}
and the Gram matrix of the basis
\begin{equation}
G_{ab} = ( \phi_a, \phi_b ) 
\end{equation}
Then $W$ becomes
\begin{equation}
\begin{aligned}
W &= \sum_{\vec{w}} \exp[ -  2\pi g (\vec{\phi}_{\text{cl}}, \vec{\phi}_{\text{cl}} )] 
= \sum_{\vec{w}} \exp[ -  2\pi g (2\pi R_c)^2 w_i^a (M^T_{n-1} M)_{ij} w_j^b G_{ab} ] \\
&= \sum_{\vec{w}} \exp[ -  2\pi g (2\pi R_c)^2 w_i^a (T_{n-1})_{ij} w_j^b G_{ab} ]
\end{aligned}
\end{equation}

We know the solution for a single hole, it is related to equation \eqref{eq:ann_sol},
\begin{equation}
  \phi_{\text{cl}} = \frac{\ln \frac{r}{r_2}}{\ln \frac{r_1}{r_2}}
\end{equation}
if we conformally map the circular hole in to the center of the disk. We take $r_1\to 0$ limit, then the solution is approaching zero away from the disk. It is hence plausible to assume that the Dirichlet boundary conditions on the other  holes only perturb the system in the small hole limit. The non-diagonal matrix element
\begin{equation}
G_{ab} = \frac{1}{2\pi} \int_{\partial A_b}   ( \nabla \phi_a )\phi_b \cdot \vec{n} \, ds
\end{equation} 
is approaching zero. In other words, we approximately have $m$ independent $\vec{w}$ vectors in this calculation. The diagonal elements are the same as in the case of annulus. It is then legitimate to use the Gaussian approximation for the sum, which gives

\begin{equation}
  -W'(1) =m  \Big[ \ln ( 4\pi R_c \sqrt{g}) - \frac{1}{2} \Big]  + \frac{1}{2} \ln \det G 
\end{equation}

It is clear that $\frac{1}{2}\ln \det G$ contains the shape dependent information about the solution. It will in general be functions of conformal invariant parameters, for example cross ratio. Since $\det G$ is a vanishing quantity, this term will be divergent. However, the divergent part will cancel those from the determinant of Laplacian in the entanglement. We have seen one such example in the cancellation of the $\ln\ln \frac{r_1}{r_2}$ in the disk EE. 

Therefore, we claim that constant contribution to the EE consists of a topological part $m\, (\ln R_c - \frac{1}{2})$  and a shape dependent factor, while the former depends on the surgery type and the later cancels possible divergences for the determinants.

\bibliography{RK_EE}

\end{document}